\documentclass{aa}
\usepackage{txfonts}
\usepackage{booktabs}
\usepackage{array}
\usepackage{graphicx}
\usepackage{natbib,twoopt}
\usepackage{upgreek}
\usepackage[breaklinks=true]{hyperref} %% to avoid \citeads line fills

\usepackage[normalem]{ulem}
\usepackage{color}

\begin{document}
        
        \title{The Apertif science verification campaign}
        \subtitle{Characteristics of polarised radio sources}
        %\subtitle{A resolved radio spectral study of the Whirlpool galaxy}
        \author{B. Adebahr\inst{1} \and A. Berger\inst{1} \and E.~A.~K.~Adams\inst{2,3} \and K.~M. Hess\inst{4,2,3} \and W.~J.~G. de Blok\inst{2,12,3} \and H. D\'enes\inst{2} \and V.~A. Moss\inst{5,6,2} \and R. Schulz\inst{2} \and J.~M. van der Hulst\inst{2} \and L. Connor\inst{13,2} \and S. Damstra\inst{2} \and B. Hut\inst{2} \and M.~V. Ivashina\inst{11} \and G.~M. Loose\inst{2} \and Y. Maan\inst{9,2} \and A. Mika\inst{2} \and H. Mulder\inst{2} \and M.~J. Norden\inst{2} \and L.~C. Oostrum\inst{2,7,10} \and E. Orr\'u\inst{2} \and M. Ruiter\inst{2} \and R. Smits\inst{2} \and W.~A. van Cappellen\inst{2} \and J. van Leeuwen\inst{2,7} \and N.~J. Vermaas\inst{2} \and D. Vohl\inst{7,2} \and J. Ziemke\inst{2,8}}
        \institute{Astronomisches Institut der Ruhr-Universit\"at Bochum (AIRUB), Universit\"atsstrasse 150, 44801 Bochum, Germany
                \and ASTRON, the Netherlands Institute for Radio Astronomy, Oude Hoogeveensedijk 4,7991 PD Dwingeloo, The Netherlands
                \and Kapteyn Astronomical Institute, University of Groningen, PO Box 800, 9700 AV Groningen, The Netherlands
                \and Instituto de Astrofísica de Andalucía (CSIC), Glorieta de la Astronomía s/n, 18008 Granada, Spain
                \and CSIRO Astronomy and Space Science, Australia Telescope National Facility, PO Box 76, Epping NSW 1710, Australia
                \and Sydney Institute for Astronomy, School of Physics, University of Sydney, Sydney, New South Wales 2006, Australia
                \and Anton Pannekoek Institute, University of Amsterdam, Postbus 94249, 1090 GE Amsterdam, The Netherlands
                \and University of Oslo Center for Information Technology, P.O. Box 1059, 0316 Oslo, Norway
                \and National Centre for Radio Astrophysics, Tata Institute of Fundamental Research, Pune 411007, Maharashtra, India
                \and Netherlands eScience Center, Science Park 140, 1098 XG, Amsterdam, The Netherlands
                \and Dept.\ of Electrical Engineering, Chalmers University of Technology, Gothenburg, Sweden
                \and Department of Astronomy, University of Cape Town, Private Bag X3, Rondebosch 7701, South Africa
                \and Cahill Center for Astronomy, California Institute of Technology, Pasadena, CA, USA}
        
        \date{-}
        
        \abstract{The characteristics of the polarised radio sky are a key ingredient in constraining evolutionary models of magnetic fields in the Universe and their role in feedback processes. The origin of the polarised emission and the characteristics of the intergalactic medium on the line of sight can be investigated using large samples of polarised sources. Ancillary infrared (IR) and optical data can be used to study the nature of the emitting objects.}{We analyse five early science datasets from the APERture Tile in Focus (Apertif) phased array feed system to verify the polarisation capabilities of Apertif in view of future larger data releases. We aim to characterise the source population of the polarised sky in the L-Band using polarised source information in combination with IR and optical data.}{We use automatic routines to generate full field-of-view Q- and U-cubes and perform Rotation Measure (RM)-Synthesis, source finding, and cross-matching with published radio, optical, and IR data to generate polarised source catalogues. All sources were inspected individually by eye for verification of their IR and optical counterparts. Spectral energy distribution (SED)-fitting routines were used to determine photometric redshifts, star-formation rates, and galaxy masses. IR colour information was used to classify sources as active galactic nuclei (AGN) or star-forming-dominated and early- or late-type.}{We surveyed an area of 56\,deg$^2$ and detected 1357 polarised source components in 1170 sources. The fraction of polarised sources is 10.57\,\% with a median fractional polarisation of $4.70\pm0.14$\,\%. We confirmed the reliability of the Apertif measurements by comparing them with polarised cross-identified NVSS sources. Average RMs of the individual fields lie within the error of the best Milky Way foreground measurements. All of our polarised sources were found to be dominated by AGN activity in the radio regime with most of them being radio-loud (79\,\%) and of the Fanaroff-Riley (FR)II class (87\,\%). The host galaxies of our polarised source sample are dominated by intermediate disc and star-forming disc galaxies. The contribution of star formation to the radio emission is on the order of a few percent for $\approx10\,\%$ of the polarised sources while for $\approx90\,\%$ it is completely dominated by the AGN. We do not see any change in fractional polarisation for different star-formation rates of the AGN host galaxies.}{The Apertif system is suitable for large-area high-sensitivity polarised sky surveys. The data products of the polarisation analysis pipeline can be used to investigate the Milky Way magnetic field on projected scales of several arcminutes as well as the origin of the polarised emission in AGN and the properties of their host galaxies.}
        
        
        \maketitle
        
        \section{Introduction}
        
        Surveys of the radio sky offer a unique opportunity to study the characteristics and evolution of galaxies in the Universe. Radio emission is capable of avoiding obscuration by dense gas, which is known to affect optical and infrared (IR) studies. Radio continuum emission traces one of the key ingredients needed for a full understanding of the physics of the Universe, namely the magnetic field. At radio wavelengths in the centimetre regime and beyond, continuum emission is dominated by gyrating relativistic electrons, which exhibit synchrotron radiation. While the total radio synchrotron emission traces the overall magnetic field, the polarised emission uncovers the degree to which it is ordered.
        
        It is known that the structure and strength of magnetic fields are important for jet collimation in active galactic nuclei (AGN) and the collapse of molecular clouds leading to star formation in galaxies (see \citet{2017SSRv..207....5R} and \citet{2012ARA&A..50...29C} and references therein). These magnetic fields are often expelled into the intergalactic space by AGN, producing powerful radio jets and/or consecutive supernova explosions forming superbubbles generated by star formation. Both processes are known to play a key role in the enrichment of intergalactic space with particles and magnetic fields \citep{2014ARA&A..52..589H}. 
        
        While AGN-dominated objects are more powerful, star-forming galaxies are more numerous, meaning that which object class influences the enrichment of the interstellar medium the most   is a matter of debate, as is the degree to which host galaxies can re-accrete the expelled material \citep{2014ARA&A..52..589H}. Therefore, understanding these processes on galactic and larger scales and their influence over the lifetime of the Universe is key to a thorough understanding of galaxy evolution.
        
        Observing large fields in the radio regime allows investigation of this context via a statistical approach. However, total power and polarised radio sky surveys are biased towards the detection of different types of objects. The radio sky at the milliJansky(mJy)-level in total intensity consists of emission from AGN and star formation with increasing dominance of the latter towards lower flux densities \citep{2018A&A...614A..47N,2018MNRAS.481.4548P}. On the other hand, the polarised radio sky is mostly dominated by AGN emission \citep{2014MNRAS.440.3113H,2014ApJ...787...99S}. Large statistical samples of radio sources in combination with information from their IR and optical emission not only help us to understand the proportions of AGN and star-forming galaxies \citep{2012MNRAS.426.3334M,2014MNRAS.440.3113H}, but also provide insight into the characteristics of their host galaxies.
        
        Up to now, all unresolved or barely resolved polarised extragalactic objects were identified as AGN. While the origin of the activity is known to be caused by supermassive black holes (SMBH), the timing and intensity of their activity are still unknown \citep{2019A&A...622A..17S,2020MNRAS.496.1706S,2021A&A...648A...9M}. While some are found to be compact and their emission confined to the host galaxies \citep{2021A&ARv..29....3O}, others exhibit megaparsec(Mpc)-scale jets \citep{1996MNRAS.279..257S}. These jets often show brightness enhancements towards their cores or lobes and are usually referred to as Fanaroff-Riley-class (FR) I and II objects \citep{1974MNRAS.167P..31F}, respectively. While polarised emission from FRII-type AGN is primarily found at the borders of their radio lobes, FRI-type AGN often show polarised emission all over their extension. FRII objects are commonly found to be more luminous and in less dense environments. The extent to which the parameters of the surrounding medium ---such as the morphology, gas composition, and gas density of the host galaxy, as well as the magnetic field distribution and gas density in the intragroup and intracluster medium surrounding the host galaxy--- are influencing the development into either an FRI or FRII source is not known and a detailed understanding of their host galaxies and their evolution is necessary \citep{2013MNRAS.430.3086G,2017MNRAS.466.4346M,2019MNRAS.488.2701M,2019MNRAS.482.5625R,2021A&A...648A.102V}.
        
        Comparing data from radio wavelengths ---where the direct synchrotron emission from AGN activity is often dominating--- with either IR or optical data ---where the prime emission mechanism is directly connected to the host galaxies--- is an important tracer of AGN activity and the influence of star formation on feedback processes. Such studies led to the discovery of the famous radio--far-infrared(FIR) correlation for star-forming galaxies \citep{1985A&A...147L...6D} or the classification of AGN into radio-loud and radio-quiet \citep{1989AJ.....98.1195K}. While more luminous AGN are often also brighter in the IR, a subclass is IR-faint \citep{2006AJ....132.2409N}. These IR-faint radio sources are known to be high-z objects, which are mostly compact and/or often highly obscured \citep{2014MNRAS.439..545C,2019MNRAS.484.1021O}. The percentage of these objects visible in faint radio surveys especially with respect to polarisation is not known. 
        Simulations and observations show that the amount of polarised sources and their emission strength is by an order of magnitude smaller than their total power counterparts \citep{2019MNRAS.482....2B}, meaning that large areas have to be surveyed with high sensitivities in order to accomplish a statistically relevant sample. Analysis in the past, mostly due to technical limitations, could only reach such sensitivities for areas of several square degrees with integration times of several tens or hundreds of hours. These observations were mainly targeted at fields where ancillary data were available to maximise the gain from the long observation times. Recent improvements in receiver and computing technology now allow simultaneous observations of large areas on the sky using phased array and phased-array feed technologies \citep{2008AIPC.1035..265V,2021PASA...38....9H,2021arXiv210914234V}. This allows the execution of blind large-area  surveys with sensitivity previously only achievable by targeted observations.
        
        APERture Tile in Focus (Apertif) is a new phased-array feed for the Westerbork Synthesis Radio Telescope (WSRT), in place on 12 of the 14 dishes. Phased-array feeds (PAFs) work by placing multiple elements in the focal plane of a dish and correlating them to generate multiple beams. Apertif combines the signal from 121 Vivaldi elements to provide 40 simultaneous beams on the sky with a combined field of view of approximately 6.6\,deg$^2$. A full overview of the Apertif system is provided by \citet{2021arXiv210914234V}.
        
        In this publication we use Apertif data of five observations taken during the science verification campaign (SVC) to demonstrate the polarisation capabilities of the system and verify the reliability of the measurements. We describe the observations in Sect. \ref{sect_observations} and the reduction of the data in Sect. \ref{sect_data_reduction}. The individual steps for the polarisation data analysis as well as an assessment of the polarisation leakage are described in Sect. \ref{sect_analysis}. Section \ref{sect_results} portrays the source statistics and compares the results with previous ones. In Sect. \ref{sect_source_characteristics} we analyse the characteristics of the polarised sources in view of the host galaxy type and the AGN- and star-formation activity. We discuss our results in Sect. \ref{Sect_discussion} and summarise in Sect. \ref{sect_summary}.
        
        In this publication we use the cosmological parameters $H_0=69.6$, $\Omega_{M}=0.286$ and $\Omega_{vac}=0.714$ from \citet{2014ApJ...794..135B}.
        
        \section{Observations}
        \label{sect_observations}
        
        The Apertif SVC was carried out between 18 March 2019 and 15 April 2019. Its aim was to verify the scientific usage of the Apertif system. Two weeks within this time period were dedicated to taking correlated imaging data including full polarisation information and calibration. The goal of these observations was scientific commissioning and finalisation of the survey strategy. During the whole period, the science teams monitored the data as it was acquired and provided immediate feedback on its quality.
        
        \begin{figure*}[htb]
                \resizebox{\hsize}{!}{\includegraphics{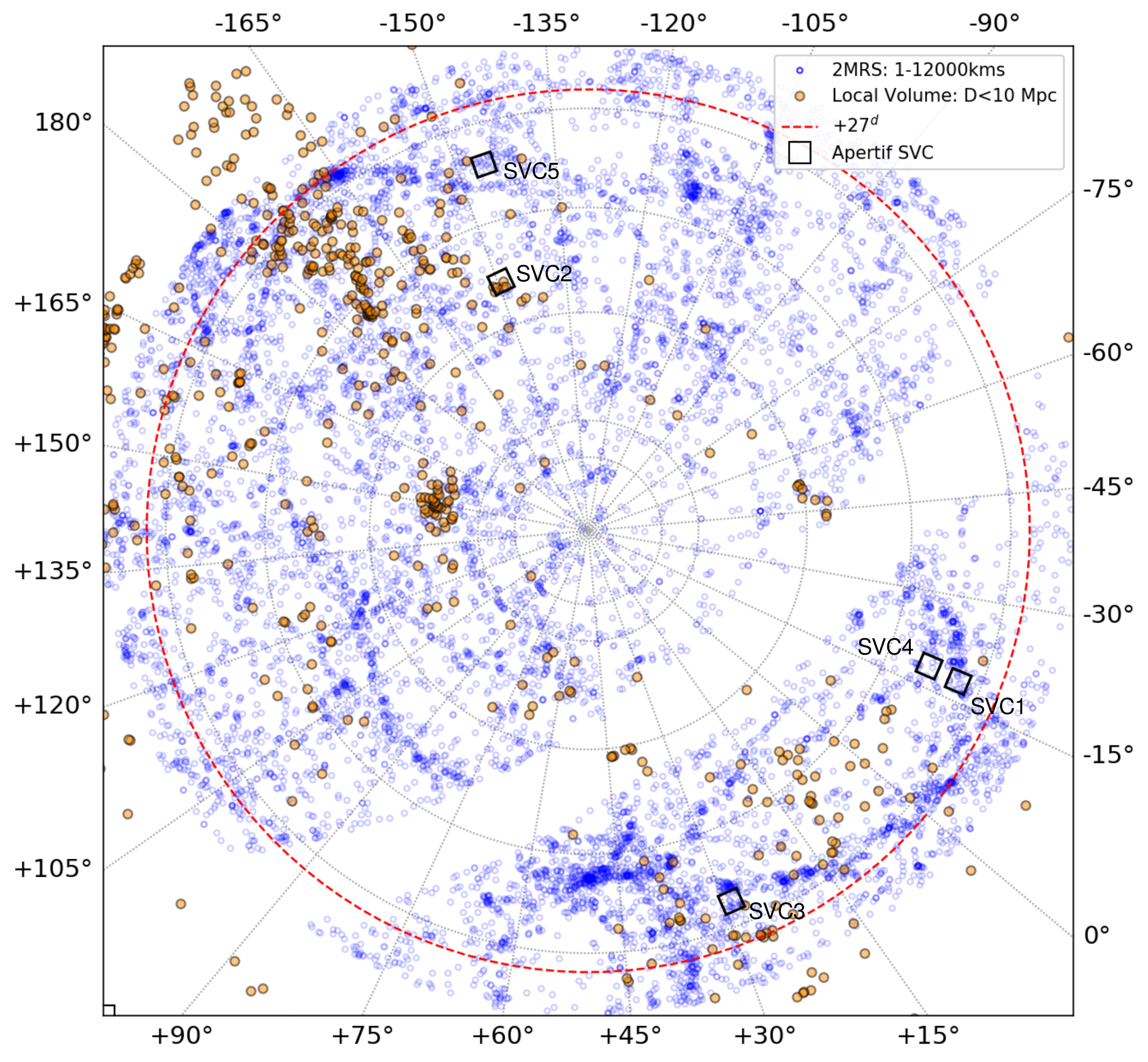}}
                \caption{Azimuthal north-pole equal-area projection showing the footprint of the SVC fields. Blue and orange circles are objects from the 2MRS catalogue \citep{2012ApJS..199...26H}  and the Local Volume survey \citep{2004AJ....127.2031K}, respectively. The red dotted line represents a declination of +27$^\circ$, which is the limit for the Apertif surveys. The five black boxes show the position and size of the SVC survey fields. The blank region represents the Milky Way.}
                \label{image_svc_footprint}
        \end{figure*}
        
        Five datasets between 8 and 11 April 2019 were recorded during this period, which are representative of the final achievable data quality of the imaging surveys. The data were publicly released through the Apertif Long Term Archive (ALTA) \footnote{\url{https://alta.astron.nl}}. Figure \ref{image_svc_footprint} shows the location of the five SVC fields.
        
        The observations occurred at a central frequency of 1370\,MHz with a bandwidth of 300\,MHz divided into 384 sub-bands with 64 channels each resulting in a frequency resolution of 12.2\,kHz. All four linear correlations (XX, YY, XY, YX) were recorded. Due to the east to west alignment of the WSRT dishes, each field was observed for 11.5\,h on source to provide sufficient (u,v) coverage. Each two observations were bracketed with four- to five-minute calibrator scans centred in each of the 40 compound beams. Alternating scans were performed on the unpolarised calibration source 3C196 and the polarised calibration source 3C138. An exception to this strategy is the last observation (field S1415+36), where only a single directly associated (e.g. bracketing) calibrator scan was carried out. Therefore, a non-bracketing calibrator scan was used from earlier in the observing run to provide full polarimetric calibration. The observed fields and calibrator information are listed in Table \ref{table:observations}.
        
        \begin{table*}[htb]
                \caption{Datasets of the SVC imaging survey fields}
                \label{table:observations}
                \centering
                \newcolumntype{0}{>{\centering\arraybackslash} m{0.8cm} }
                \newcolumntype{1}{>{\centering\arraybackslash} m{1.5cm} }
                \newcolumntype{2}{>{\centering\arraybackslash} m{1.5cm} }
                \newcolumntype{3}{>{\centering\arraybackslash} m{0.8cm} }
                \newcolumntype{4}{>{\centering\arraybackslash} m{1.0cm} }
                \newcolumntype{5}{>{\centering\arraybackslash} m{2.2cm} }
                \newcolumntype{6}{>{\centering\arraybackslash} m{0.8cm} }
                \newcolumntype{7}{>{\centering\arraybackslash} m{1.0cm} }
                \newcolumntype{8}{>{\centering\arraybackslash} m{2.2cm} }
                \newcolumntype{9}{>{\centering\arraybackslash} m{0.8cm} }
                \begin{tabular}{@{} 0 1 2 3 | 4 5 6 | 7 8 9 @{}}
                        \toprule
                        \toprule
                        \multicolumn{4}{c|}{Target field} &
                        \multicolumn{3}{c|}{Flux calibrator} &
                        \multicolumn{3}{c}{Polarisation calibrator} \\
                        Field & Name & Task ID &  t$_{obs}$ (min) & Name & Obs IDs &  t$_{obs}$ (min) & Name & Task IDs & t$_{obs}$ (min) \\
                        \midrule
                        SVC1 & S2248+33\tablefootmark{a} & 190409015 & 690 & 3C196  & 190408125-150  190409001-014\tablefootmark{a} & 5 & 3C138 & 190409016-055 & 4 \\
                        SVC2 & M1403+53 & 190409056 & 690 & 3C196 & 190410002-041 & 5 & 3C138 & 190409016-055 & 4 \\
                        SVC3 & M0155+33 & 190410001 & 690 & 3C196 & 190410002-041 & 5 & 3C138 & 190409016-055 & 4  \\
                        SVC4 & S2246+38 & 190411001 & 690 & 3C196 & 190410002-041 & 5 & 3C138 & 190411002-041 & 4  \\
                        SVC5 & S1415+36 & 190411042 & 690 & 3C196\tablefootmark{b} & 190410002-041\tablefootmark{b} & 5 & 3C138 & 190411002-041 & 4 \\
                        \bottomrule
                \end{tabular}
                \tablefoot{
                        \tablefootmark{a}{Beams 31-39 failed (190409006-14 not on source) }
                        \tablefootmark{b}{Non-bracketing flux calibrator used due to failure of observing session}
                }
        \end{table*}
        
        \section{Data reduction}
        \label{sect_data_reduction}
        
        During the SVC period, the imaging data were automatically run through the Apertif imaging pipeline (Apercal) \citep{2022A&C....3800514A}. In the following, we want to describe the main calibration and imaging steps performed by Apercal which are relevant for the generation of the continuum and polarisation data products used in this publication. Apercal is publicly available on github\footnote{\url{https://github.com/apertif/apercal/releases}}. For the SVC processing release version 2.4, `SVC-Reprocessing' was used.
        
        The data were acquired from ALTA. Early science data suffer from delay issues due to an incomplete correction of the phase tracking for all beams. For survey data, this correction is automatically applied before data are ingested into ALTA. Specifically for the SVC data, an offline correction to the fringe-stopping was applied immediately after retrieving the data. To mitigate the influence of strong RFI and enhance the performance of automatic flagging routines, only the frequency range between 1291.8\, and 1441.8\,MHz was used for further processing resulting in an effective bandwidth of 150\,MHz.
        
        Flagging was executed in three steps: First, continuous problematic data ranges were flagged. This included the first, central, and last channels of each subband as well as flagging for shadowing of dishes. As a next step, the data were carefully manually inspected for additional RFI or system issues. For all fields, beams 16 and 18 were not processed as an inspection of their phases revealed a problematic sub-band-based behaviour. For beam 01, RT9 was also flagged for all observations as it showed variations and jumps in the phases every 20 sub-bands. The first target field (S2248+33) also had several additional flags applied. Beams 31-39 were not processed as the flux calibrator observations were not successful. In addition, RTD was flagged for beams 0-5; high temperatures in the electronics cabinet for this dish resulted in some components shutting down for that time period, meaning corrupted data were recorded. For the last target dataset (S1415+36), RTC was flagged due to residual delay issues. Finally, we note that channels (of the full dataset) 10752--12287 (1380--1400 MHz) for the YY polarisation of RT5 are automatically flagged by the system for the SVC data; this is a result of the correlator not receiving the data.
        
        The last flagging step was performed automatically using aoflagger \citep{2012A&A...539A..95O} with a flagging strategy specifically tailored to the parameters of the Apertif data (see Offringa et al. in prep. and \citet{2010MNRAS.405..155O} for further details). The strategy uses a low-pass filter in combination with the Sumthreshold and Scale Invariant Rank operations to detect spurious radio frequency interference. 
        
        Residual delays, complex gains, bandpasses, and polarisation leakage solutions were derived using the flux calibrator datasets. The polarisation angle and time-dependent phase offsets between the feeds were derived using the polarised calibrator dataset. For all steps, the Common Astronomical Software Application (CASA) \citep{2007ASPC..376..127M} was used. All solutions were then applied to the appropriate target datasets.
        
        Further processing was performed using the Multichannel Image Reconstruction Image Analysis and Display software (MIRIAD) \citep{1995ASPC...77..433S}. The data were first converted to the MIRIAD format and then averaged in frequency by a factor of 64, so that each individual channel encompasses the data of one sub-band. Self-calibration was then performed for each individual target beam dataset. This step consisted of three substeps. First, a frequency-dependent model of the sky was generated using a combination of the catalogues of the NRAO VLA Sky Survey (NVSS) \citep{1998AJ....115.1693C}, The Faint Images of the Radio Sky at Twenty Centimeters (FIRST) Survey \citep{1995ApJ...450..559B}, and the Westerbork Northern Sky Survey (WENSS) \citep{1997A&AS..124..259R}. This parametric self-calibration step was followed by several iterations of phase self-calibration, and, in cases with sufficient signal-to-noise ratio (S/N), one iteration of amplitude self-calibration. For each self-calibration cycle, the solution interval was decreased and the (u,v)-range of data to include was increased. The specific details of the self-calibraton routines are described in \citet{2022A&C....3800514A}.
        
        The final self-calibrated data were used to produce a multi-frequency synthesis Stokes I total power image. We cleaned the resulting image down to the $1\sigma$ level within masks generated with a $5\sigma$ threshold, where $\sigma$ is the calculated theoretical noise of the observation. In addition to the total power image, we produced Stokes Q and U image cubes. For easier handling, the Q- and U-data were imaged in chunks of 6.25\,MHz bandwidth resulting in 24 final images for each Stokes parameter over the total bandwidth of 150\,MHz. Cleaning of the individual images was performed using the masks generated during the Stokes I imaging and cleaning down to the $1\sigma$ level of the noise in the individual Stokes Q- and U-images. Stokes V images were created using multi-frequency synthesis and cleaning down to the $1\sigma$ level of the noise in the image using the same masks as before.
        
        We created mosaics of all compound beams of the five different pointings in Stokes I, Q, and U. For all mosaics, the central Beam (00) was excluded. This beam delivers mostly redundant information and would therefore add correlated noise to the data. For Stokes I, all successfully calibrated multi-frequency images were used. Strong image artefacts or very high noise values of individual images within a single Q or U cube can corrupt the entire polarisation analysis. Therefore, we discarded all single image planes of individual beams with noise values of $>300\,\upmu$Jy/beam. For further polarisation analysis, all images of the same observation over the entire frequency range need the same synthesised beam size. In order to mitigate the effect of large final beams caused by individual images where data of long baselines are missing, all images within a cube with $\theta_{maj}>35''$ or $\theta_{min}>20.5''$ were discarded, where $\theta_{maj}$ and $\theta_{min}$ are the FWHM of the major and minor axis of the synthesised beam, respectively. To ensure constant parameters for the polarisation analysis over the field of view of a single pointing, the same frequency and spatial coverage has to be given for the whole Stokes Q and U mosaic cubes. We therefore discarded images of entire beams or all images at frequencies where two or more images were missing. This ensures that no images containing only NaN values enter the polarisation analysis. 
        
        \begin{table}[htb]
                \centering
                \caption{Accepted images for the SVC mosaicking routines.} 
                \label{table:mosaics}
                \begin{tabular}{lrrr}
                        \toprule
                        \toprule
                        Field & $N_{c}$ & $N_{p,f}$ & $N_{p,b}$ \\
                        \midrule
                        SVC1 & 28 & 22 & 28 \\
                        SVC2 & 37 & 23 & 31 \\
                        SVC3 & 37 & 23 & 37 \\
                        SVC4 & 37 & 22 & 32 \\
                        SVC5 & 37 & 22 & 36 \\
                        \bottomrule
                \end{tabular}
                \tablefoot{$N_c$ is the number of accepted continuum images and $N_{p,f}$ and $N_{p,b}$ are the number of individual frequency and beam images accepted for generating the Stokes Q- and U-image cube mosaics, respectively. The maximum possible numbers would be $N_c=39$, $N_{p,f}=24$ and $N_{p,b}=39$ }
        \end{table}
        
        We then convolved all accepted Stokes-Q and -U images  to the smallest common synthesised beam and primary beam corrected them using Gaussian regression models (Kutkin et al. in prep.). The images were then clipped at the 5\% level of the primary beam response and combined using an inverse square weighting of the noise in the individual images. 
        
        \section{Polarisation data analysis}
        \label{sect_analysis}
        
        In the following we present the strategy and software routines\footnote{\url{https://github.com/adebahr/aperpol}} to generate polarised source catalogues from Apertif imaging data in a semi-automatic way. The SVC datasets serve as a test-bed for the analysis of the polarisation data of the whole Apertif survey.
        
        \subsection{Rotation measure synthesis}
        
        Stokes-Q and -U fluxes from astronomical sources exhibit a sinusoidal dependence of the square of the observed wavelength. Depending on the value of the rotation measure (RM) of the received signal, this can lead to depolarisation within the observing band. To mitigate this effect, we performed RM synthesis \citep{2005A&A...441.1217B} on the final Stokes Q- and U-mosaic cubes to generate Faraday Q- and U-cubes. The resolution in Faraday space $\delta\Phi$ is approximated by
        \begin{equation}
        \delta\Phi\approx\frac{2\sqrt{3}}{\Delta\lambda^2},
        \end{equation}
        the maximum Faraday scale to which sensitivity has dropped to 50\,\% is 
        \begin{equation}
        \Phi_{ms}\approx\frac{\pi}{\lambda_{min}^2}
        ,\end{equation}
        and the maximum Faraday depth with more than 50\,\% sensitivity is
        \begin{equation}
        ||\Phi_{max}||\approx\frac{\sqrt{3}}{\delta\lambda^2},
        \end{equation}
        where $\delta\lambda^2$ is the channel width, $\Delta\lambda^2$ the width of the $\lambda^2$ distribution, and $\lambda_{min}^2$ is the shortest wavelength squared. We list the resulting parameters for our setup for all five of our fields in Table \ref{table:RM_params}.
        
        \begin{table}[htb]
                \centering
                \caption{RM synthesis parameters for all five SVC fields.} 
                \label{table:RM_params}
                \begin{tabular}{lccc}
                        \toprule
                        \toprule
                        Field & $\delta\Phi$ & $\Phi_{ms}$ & $\Phi_{max}$ \\
                        & [rad/m$ 2$] & [rad/m$ 2$] & [rad/m$ 2$] \\
                        \midrule
                        SVC1 & 378.33 & 72.34 & 3990.65 \\
                        SVC2 & 353.60 & 71.72 & 3909.46 \\
                        SVC3 & 353.60 & 71.72 & 3909.46 \\
                        SVC4 & 378.33 & 72.34 & 3990.65 \\
                        SVC5 & 378.33 & 72.34 & 3990.65 \\
                        \bottomrule
                \end{tabular}
                \tablefoot{$\delta\Phi$ is the resolution in Faraday space, $\Phi_{ms}$ the maximum observable Faraday scale, and $\Phi_{max}$ the maximum observable Faraday depth.}
        \end{table}
        
\begin{figure}[h]
                \includegraphics{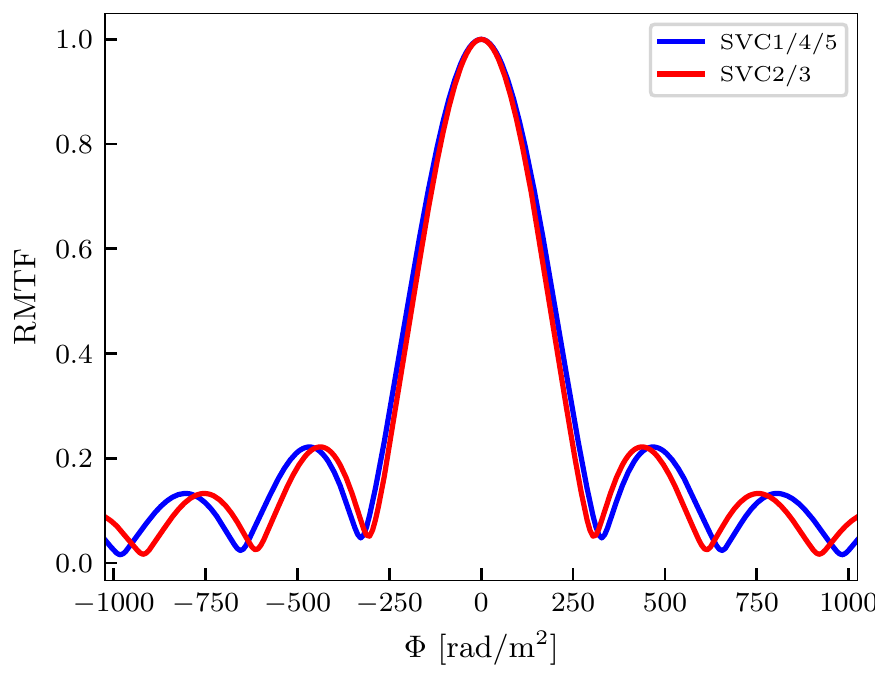}
                \caption{Rotation-measure transfer function for the SVC Faraday cubes for our sampled Faraday range of -1024\,rad/m$^2\leq\Phi\leq$ 1024\,rad/m$^2$.}
                \label{plot_RMTF}
        \end{figure}

        We sampled the Faraday axis in the range of -1024\,rad/m$^2\leq\Phi\leq$ 1024\,rad/m$^2$ with a sampling interval of 8\,rad/m$^2$ resulting in Q- and U-Faraday cubes of 257 planes each. The resulting rotation-measure transfer functions (RMTFs) are shown in Fig. \ref{plot_RMTF}. The first sidelobes of the RMTF are located at $\sim$500\,rad/m$^2$ and on a level of $\sim20\%$. We want to note that the small discrepancy in the two functions and the RM-synthesis parameters (Table \ref{table:RM_params}) of fields SVC1/4/5 and SVC2/3 originates from the slightly different frequency coverage of the input images.
        
        \subsection{Faraday cube analysis}
        
        We determined polarised intensity (PI) and RM values from the resulting Faraday cubes. First, the absolute of the complex polarisation vector Q+iU is calculated, which results in a polarised intensity (PI) image cube. The maximum of the PI along the Faraday axis represents the linear polarised flux of the main polarised component along the axis. The position of this peak is the RM value. For optimal determination of these values and to overcome the limited sampling of the Faraday axis of 8\,rad/m$^2$, we search for the highest value along the Faraday axis for all pixels in RA- and DEC directions in our PI image cube. We then fit a one-dimensional parabola in Faraday space to this value and the two neighbouring values. The resulting PI- and RM values are saved to two-dimensional PI- and RM maps. We want to note that this analysis technique is only sensitive to the brightest component in Faraday space. Manual inspection of the Faraday cubes would be needed for analysis of secondary or higher order components. 
        
        \begin{figure*}[htb]
                \begin{center}
                        \resizebox{\hsize}{!}{\includegraphics{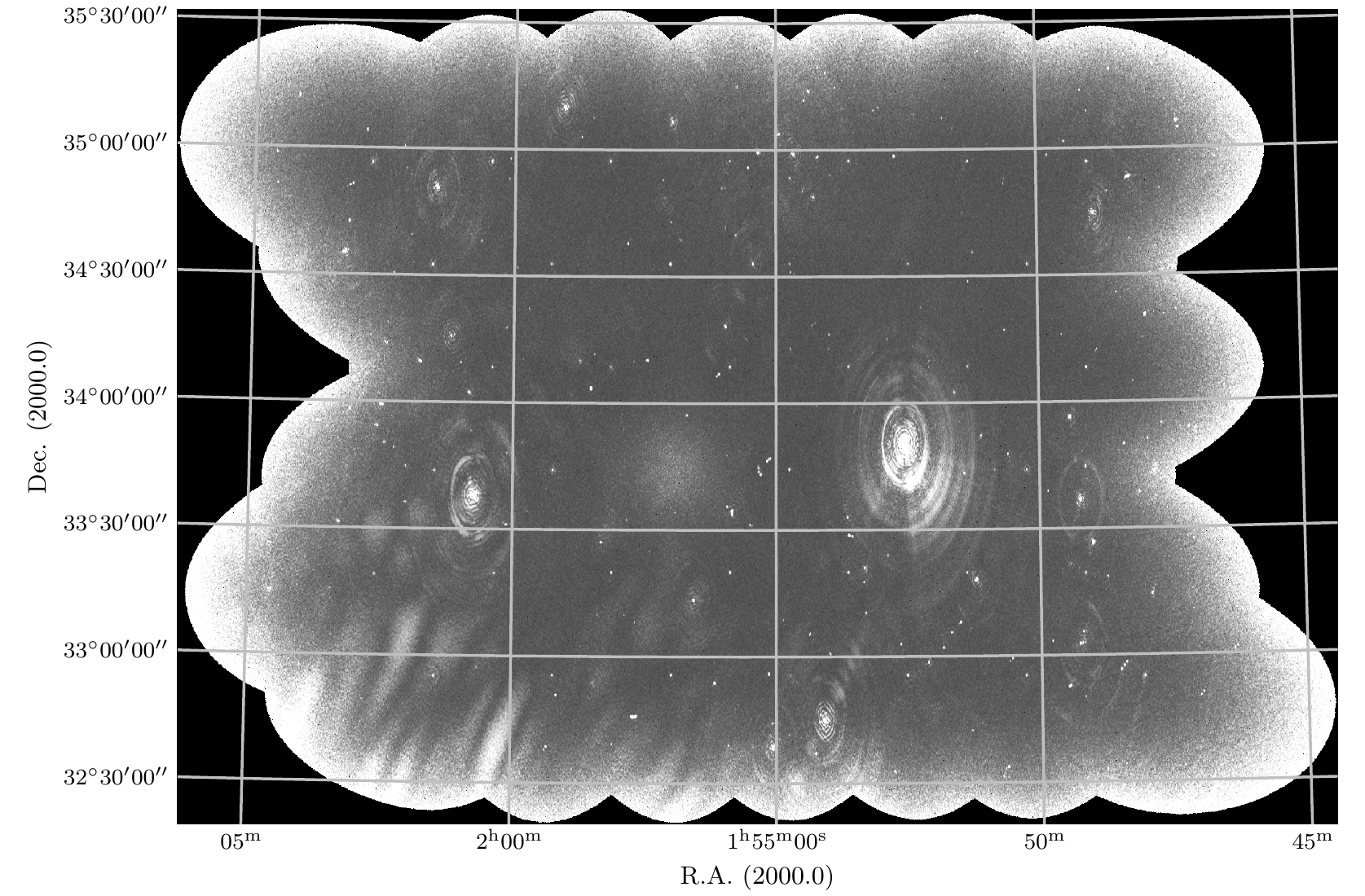}}
                        \caption{Polarised intensity image of the field M0155+33 (SVC3). The noise in emission-free regions is 14\,$\upmu$Jy/beam. The synthesised beam size is $29.2''\times16.1''$. Beams 16 and 18 are missing, which appears as increased noise towards the southeast and southwest of the central position. Several sources show artefacts around their positions, which can be traced back to direction-dependent calibrations issues. The diffuse polarised emission in the southeast originates from Galactic foreground emission. Polarised intensity images for the other SVC fields are provided in Appendix \ref{appendix:piimages}.}
                        \label{image_PI_SVC}
                \end{center}
        \end{figure*}
        
        \begin{table*}[htb]
                \centering
                \caption{Parameters of the SVC mosaics.} 
                \label{table:SVC_params}
                \begin{tabular}{lccccccc}
                        \toprule
                        \toprule
                        Field & Date of observation & RA\tablefootmark{a} & DEC\tablefootmark{a} & $\sigma_{PI}$\tablefootmark{b} & FWHM$_{PI}$ & $\sigma_{TP}$\tablefootmark{b} & FWHM$_{TP}$ \\
                        & [dd-mm-yyyy] & [hh:mm:ss] & [dd:mm:ss] & [$\upmu$Jy/beam] & [''] & [$\upmu$Jy/beam] & [''] \\
                        \midrule
                        SVC1 & 09-04-2019 & 22:48:39 & +33:56:40 & 14 & $32.1\times18.0$ & 35 & $27.2\times12.1$ \\
                        SVC2 & 09-04-2019 & 14:03:24 & +53:24:05 & 13 & $19.3\times14.6$ & 43 & $18.6\times12.8$ \\
                        SVC3 & 10-04-2019 & 01:55:19 & +33:56:40 & 14 & $29.2\times16.1$ & 42 & $26.2\times11.7$ \\
                        SVC4 & 11-04-2019 & 22:46:24 & +38:48:32 & 16 & $24.8\times14.8$ & 39 & $23.7\times12.0$ \\
                        SVC5 & 11-04-2019 & 14:15:53 & +36:22:36 & 16 & $31.2\times17.7$ & 46 & $25.5\times11.9$ \\
                        \bottomrule
                \end{tabular}
                \tablefoot{
                        \tablefoottext{a}{Coordinates correspond to the telescope pointing positions.}
                        \tablefoottext{b}{$\sigma_{PI}$ and $\sigma_{TP}$ are the noise levels of the polarised intensity and total power images, respectively, which were derived by determining the lowest quartile of the pixel values of the corresponding rms maps from the PyBDSF source finding.}
                }
        \end{table*}
        
        \subsection{Source finding}
        \label{subsection_source_finding}
        
        We performed source finding in the final PI maps using the Python Blob Detector and Source Finder (PyBDSF) \citep{2015ascl.soft02007M}. Our final PI maps show artefacts around bright sources. These are caused by directional calibration issues in combination with stacking of uncleaned sidelobes in the individual Q- and U-images. To mitigate the influence of these artefacts on our source-detection analysis, we used an rms box with an adaptive threshold of $10.0\sigma$. The island threshold was set to $5.0\sigma$ and the pixel threshold to $6.25\sigma$. The kernel filter parameter \textit{rms\_box} was set to (60,20). To reduce false detections due to artefacts in the vicinity of strong sources even further, we enabled the \textit{rms\_box\_bright} parameter with values of (20,8).
        
        A source-detection mask and a source catalogue were generated for each field. In order to calculate the fractional polarisation (FP) values of sources and generate a total power (TP) catalogue for cross-matching in later stages of the analysis, we performed another PyBDSF run on the Stokes I TP image. Here, we used slightly different parameters because of the inherently different distribution of sources and noise characteristics compared to the PI maps. An adaptive rms box was enabled as well. We used thresholds of $4.0\sigma$ and $5.0\sigma$ for islands and pixels, respectively. The parameter \textit{rms\_box} was set to (100,10) and \textit{rms\_box\_bright} to (20,2). In addition we forced the mean map to be zero and fitted a rank-1 spline to the background.
        
         We convolved the Stokes I total power image to the resolution of the PI-map when calculating the FP maps. We then divided the PI map by the TP map. In order to limit the PI, RM, and FP maps to regions of sufficient S/N, we used the image masks generated by PyBDSF for the PI source finding to blank any pixels outside of the masked regions for these images.
        
        Our catalogues contain one entry for each detected component. Sources with multiple components are associated by the same source ID. PI- and TP flux densities as well as FP values are given individually for each component and for the integrated fluxes of the sources. Calculating average RMs for sources, especially in a resolved case, is difficult because of the strongly varying values close to the borders of masks originating from lower S/Ns. Therefore, we only include RM values for each individual component. These values were determined using the RA- and DEC coordinates of the component in the PI catalogue. For a description of the columns in the catalogue\footnote{The catalogue data is available at \url{https://vo.astron.nl/}}, see Appendix \ref{appendix_source_catalogue}.
        
        \subsection{Cross-matching}
        \label{subsect_cross_matching}
        
        The RA- and DEC coordinates of each source in the PI catalogue were used for cross-matching with our generated TP catalogues, the NVSS RM catalogue\citep{2009ApJ...702.1230T}, the All Wide-field Infrared Survey Explorer (WISE) database \citep{2014yCat.2328....0C}, and the Sloan Digital Sky Survey (SDSS) DR16 database \citep{2020ApJS..249....3A}. While full coverage of the SVC footprint is provided by the NVSS and AllWISE databases, the SDSS coverage is 65.52\,\% (36.94\,deg$^2$ out of 56.38\,deg$^2$). To acquire the best possible results and keep false cross-matching rates to a minimum, we proceeded in the following way.
        
        An aliasing problem in the Apertif correlator is known to produce fake sources at the central position of each compound beam. We made sure to exclude these false detections from the catalogues by removing any detected TP- or PI source within the radius of the FWHM of the synthesised beam in the central observed pointing positions of each compound beam from our databases.
        
        For each PI source, the closest TP source located within the major radius of the synthesised beam of the PI image was associated as its counterpart. If a TP source was found, the AllWISE database\footnote{\url{https://wise2.ipac.caltech.edu/docs/release/allwise/}} was queried. As polarised emission is often not coincident with the central position of an object, especially for extended objects, we used the coordinates of the total power cross-matched source. An AllWISE cross-match was identified if an entry could be located within the major radius of the synthesised beam of the TP image. The position, brightness, and S/N in all four WISE bands (3.4\,$\upmu$m, 4.6\,$\upmu$m, 12\,$\upmu$m, 22\,$\upmu$m) and their errors were acquired and added to our database. Due to the better correlation of IR data to optical data, we cross-matched our objects with the SDSS DR16 database\footnote{\url{https://www.sdss.org/dr16/}} using the WISE coordinates. This strategy has the advantage of producing much less false cross-matching results than a direct cross-match using our radio continuum coordinates to SDSS. The closest SDSS match was identified within the WISE resolution of 3.6''. If an object was found, its brightness information in all five SDSS bands (u, g, r, i, z) was added to our database including the redshift information, where available.
        
        For further analysis and verification of our results, we cross-matched all polarised sources with the NVSS RM catalogue. Sources within the radius of the synthesised beam of the PI image were checked. In case of a match, the PI-, FP-, and RM NVSS values were added to our database.
        
        \subsection{Ancillary data}
        
        Ancillary data products generated during the analysis include images of all four AllWISE bands and the SDSS g-band. Individual images for the entire extent of each PI-image mosaic were downloaded and mosaicked using the astropy affiliated Python wrapper MontagePy\footnote{\url{https://montage-wrapper.readthedocs.io/en/v0.9.5/}} for the  Montage Astronomical Image Mosaic Engine. 
        
        Mosaicking of the images was performed by first reprojecting them to the PI image mosaic, adjusting and correcting the background of all images within a pointing, and finally co-adding them to produce a single image spanning the extent of each PI mosaic of one pointing. The final images are especially useful for the manual inspection of sources and their cross-matches in later stages of the analysis (see Sect. \ref{subsection_catalogue}).
        
        \subsection{Catalogue generation}
        \label{subsection_catalogue}
        
\begin{figure*}[htb]
                \resizebox{0.94\textwidth}{!}{\includegraphics{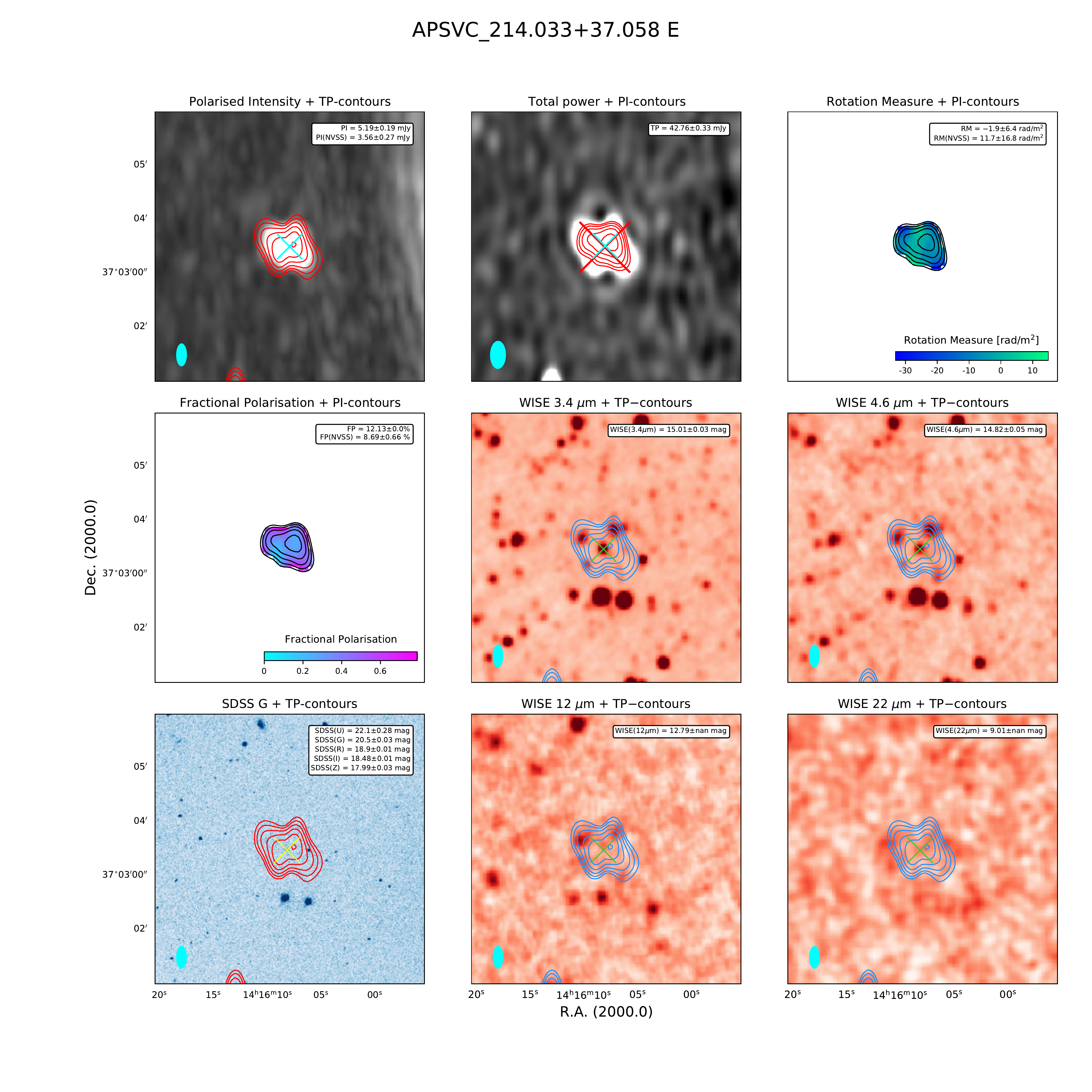}}
                \caption{Example catalogue pdf document of the source APSVC\,214.033+37.058 detected in field S1415+36 (SVC5). The source ID and the source code (S for point-like, E for extended) are given in the caption. The images show the following maps and contours: \textit{Top left:} Polarised intensity and total power continuum contours. \textit{Top centre:} Total power continuum and polarised intensity contours. \textit{Top right:} Rotation measures and polarised intensity contours. \textit{Centre left:} Fractional polarisation and polarised intensity contours. \textit{Centre:} AllWISE 3.4\,$\upmu$m and total power continuum contours. \textit{Centre right:} AllWISE 4.6\,$\upmu$m and total power continuum contours.  \textit{Bottom left:} SDSS and total power continuum contours.  \textit{Bottom centre:} AllWISE 12\,$\upmu$m and total power continuum contours.  \textit{Bottom right:} AllWISE 22\,$\upmu$m and total power continuum contours. All PI contours start at the 6.25\,$\sigma$-level of the local noise and increase in powers of $\sqrt{2}$. TP contours start at the 7\,$\sigma$-level of the local noise and increase in powers of two. Cyan crosses in the top left and top central images show the central source positions of the PI and TP sources, respectively. A red cross in the top central images shows the position of a source identified in NVSS, if available. Green and yellow crosses in the WISE and SDSS images mark the positions of the cross-matched sources of these catalogues, respectively. Values measured from our data and extracted ones from the cross-matched catalogues are always shown in the top right corner of the relevant images.}
                \label{image_catalogue}
        \end{figure*}

        We generated different overlays (see Fig. \ref{image_catalogue}) and saved them into a portable document format (pdf) file to inspect each source in the PI catalogue individually. The association of extended and complex sources with each other as well as cross-matches with their TP-, AllWISE-, and SDSS counterparts are checked for obvious errors. Interactive functions in the analysis pipeline allow the removal of sources from the PI catalogue, the splitting of complex sources into components, the combination of source components into a single source, and manually searching for TP-, AllWISE-, and SDSS-counterparts at given positions.
        
        Sources with no TP counterpart or those showing obvious artefacts were discarded. Individual components are combined into a single source, if they are obviously originating from the same source. This is often the case for resolved FR-class objects, where the core and hotspot or jet regions appear disconnected. AllWISE- and SDSS counterparts were removed for incorrectly identified counterparts. Often these sources are at the edge of the search radius or the central position of diffuse and/or resolved sources could not easily be calculated correctly due to their complexity. In the latter case, the AllWISE and SDSS catalogues were manually checked for the central positions of these sources   which were identified by eye. 
        
        \subsection{Polarisation leakage}
        
        To verify the reliability of the identified source catalogue, we assess the characteristics of the instrumental leakage of the Apertif system. Stokes V represents the circular polarisation, which is a very rare phenomenon in astronomical objects. \citet{kups6967} found a median degree of circular polarisation in blazar sources of 0.4\,\%. A targeted survey of 150 polarised AGN sources by \citet{2018A&A...609A..68M} revealed 10 circular polarised sources with circular polarisation degrees of approximately 0.2\,\%. \citet{2002ApJ...578L.103B} detected circular polarisation degrees around 0.1\,\% for nearby low-luminosity AGNs. Similar values were found by \citet{2000MNRAS.319..484R} and \citet{2001ApJ...556..113H} for a sample of 12 and 11 AGN sources, respectively. We highlight the fact that all of the above observations were targeted towards sources where a significant fraction of circular polarisation was expected. Therefore, we can assume that the large majority of sources with a Stokes V detection for our Apertif data are dominated by instrumental leakage.
        
        In order to set a limit where instrumental polarisation is dominant for our sources, we measure the amount of fractional circular polarisation dependent on the beam response. For this purpose, we performed a source detection using PyBDSF on each individual Stokes I and V image. Since PyBDSF is only able to detect positive values and Stokes V emission can exhibit both signs, we performed the source finding twice, once on the original image and a second time on the inverted one.
        
        The two Stokes V source catalogues were combined for each beam and all detected Stokes V sources were then cross-matched with the appropriate Stokes I sources of the inspected beam. The response of the primary beam for these sources was calculated by extracting the value at the position relative to the pointing centre from the Gaussian regression beam models (Kutkin et al. in prep.).
        
        \begin{figure}[htb]
                \includegraphics{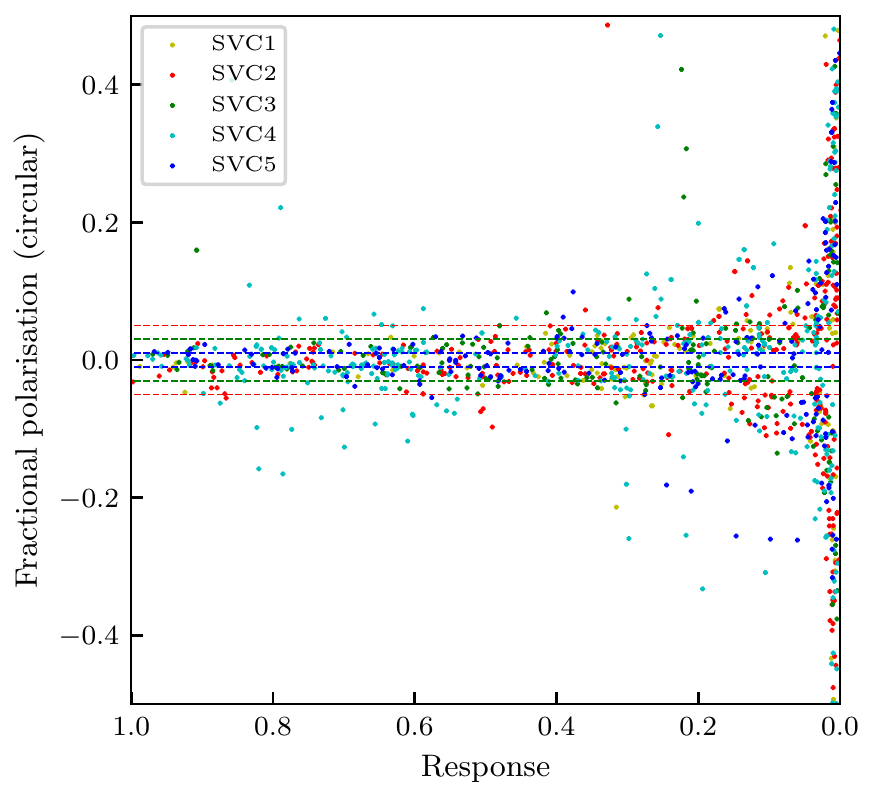}
                \caption{Primary beam responses of the detected sources in Stokes V vs. their fractional circular polarisation. The different colours mark the fields in which the sources were detected. The blue, green, and red dashed lines represent the 1\%, 2\%, and 3\% fractional polarisation levels, respectively.}
                \label{plot_leakage_response}
        \end{figure}
        
        Figure \ref{plot_leakage_response} shows the primary beam response of the detected circularly polarised sources vs. their fractional circular polarisation. The fractional polarisation stays constant at the level of about 1\% up to a primary beam response of 0.3. We do not see a specific bias towards a certain SVC field, and so we conclude that there was a constant leakage during the whole SVC campaign for our observations. Our individual beams overlap at an approximate primary beam response of 0.5. Therefore, we use a lower limit of 1\% for the fractional polarisation of our polarised sources to include them in any further steps in our analysis. 
        
        \section{Results}
        \label{sect_results}
        
        We now want to verify the reliability of our source detection by first comparing source densities and their median FP values with published works and secondly comparing the RM- and FP values of individual sources with their archival NVSS values. In addition, we analyse the usability of our RM measurements for Milky Way foreground studies. For all further analyses, we combined the data from all five survey fields to enhance our sample statistics.
        
        \subsection{Source statistics}
        \label{subsect_source_statistics}
        
        Our images cover an area of 56.38\,deg$^2$, in which we detected 12834 source components in total emission and 1357 polarised source components. This results in a fraction of 10.6\,\% of source components being polarised. \citet{2007ApJ...666..201T}, \citet{2010ApJ...714.1689G}, \citet{2014MNRAS.440.3113H}, \citet{2014ApJ...785...45R}, and \citet{2021arXiv210702492B} found fractions of 10.6\,\%, 14.2\,\%, 5.9\,\%, 2.6\,\%, and 8.8\,\%, respectively, and so our values are consistent with the previously reported ones.
        
        The polarised source density for our imaged area corresponds to $\sim21$ per deg$^2$ at an approximate noise level of 15$\upmu$Jy/beam (see Table \ref{table:SVC_params}). Several other observations at the same frequency with similar noise levels found consistent values. \citet{2014MNRAS.441.2555H} observed between 16 and 23 polarised sources per deg$^2$ at a noise level of 25$\upmu$Jy/beam. \citet{2021arXiv210702492B} found 23 polarised sources per deg$^2$ at noise levels of 7$\upmu$Jy/beam. \citet{2014ApJ...785...45R} predicted $35\pm10$ polarised sources per deg$^2$ based on their deep observations of the GOODS-N field for a flux regime of 50$\upmu$Jy/beam.
        
        \begin{table*}[htb]
                \centering
                \caption{Source statistics for the SVC data.} 
                \label{table:SVC_source_statistics}
                \begin{tabular}{ccccccccccccc}
                        \toprule
                        \toprule
                        Field & Area & N$_{TP,c}$ & N$_{PI,c}$ & P$_F$ & N$_{PI}$ & N$_{PI,S}$ & N$_{PI,E}$ & P$_{F,S}$ & P$_{F,E}$ & $\widetilde{FP}$ & $\widetilde{FP_S}$ & $\widetilde{FP_E}$ \\
                        & [deg$^2$] & & & [\%] & & & & [\%] & [\%] & [\%] & [\%] & [\%] \\
                        \midrule
                        SVC1 & 9.49 & 1961 & 199 & 10.15 & 178 & 126 & 52 & 70.80 & 29.20 & 4.79$\pm$0.34 & 4.56$\pm$0.38 & 5.59$\pm$0.67 \\
                        SVC2 & 10.67 & 3126 & 296 & 9.47 & 254 & 161 & 93 & 60.38 & 36.62 & 4.28$\pm$0.30 & 3.59$\pm$0.35 & 5.33$\pm$0.54 \\
                        SVC3 & 12.26 & 2322 & 299 & 12.88 & 254 & 166 & 88 & 65.34 & 34.66 & 5.18$\pm$0.29 & 4.70$\pm$0.39 & 5.70$\pm$0.41 \\
                        SVC4 & 11.72 & 3030 & 299 & 9.87 & 254 & 150 & 104 & 59.06 & 40.94 & 4.92$\pm$0.29 & 4.85$\pm$0.38 & 4.97$\pm$0.46 \\
                        SVC5 & 12.22 & 2395 & 264 & 11.02 & 230 & 175 & 55 & 76.06 & 23.94 & 4.51$\pm$0.33 & 3.91$\pm$0.36 & 6.77$\pm$0.70 \\
                        All & 56.38 & 12834 & 1357 & 10.57 & 1170 & 778 & 392 & 66.50 & 33.50 & 4.70$\pm$0.14 & 4.33$\pm$0.17 & 5.47$\pm$0.24 \\
                        \bottomrule
                \end{tabular}
                \tablefoot{Area is the imaged area of the entire mosaics down to the 5\,\% level, N$_{TP,c}$ and N$_{PI,c}$ are the number of total power and polarised components, respectively. P$_F$ is the percentage of polarised source components. N$_{PI}$, N$_{PI,S}$, and N$_{PI,E}$ are the number of all polarised sources, unresolved polarised sources, and resolved polarised sources, respectively. P$_{F,S}$ and P$_{F,E}$ are the percentages of unresolved and resolved polarised sources. $\widetilde{FP}$, $\widetilde{FP_S}$, and $\widetilde{FP_E}$ are the median fractional polarisations of all polarised sources, unresolved polarised sources, and resolved polarised sources, respectively.}
        \end{table*}
        
        Of the 1357 polarised source components, 1170 individual polarised sources were identified. Of these, 778 were classified as unresolved and 392 as resolved, which leads to fractions of 66.5\,\% and 33.5\,\%, respectively. The median fractional polarisation of our complete sample is $4.70\pm0.14$\,\%. Studies of individual fields with areas covering several square degrees by \citet{2014MNRAS.440.3113H}, \citet{2010ApJ...714.1689G}, \citet{2007ApJ...666..201T}, and \citet{2021arXiv210702492B} at the same wavelength report values of 6.2\,\%, 14.1\,\%, 10.6\,\%, and 5.4\,\%, respectively.  \citet{2014ApJ...787...99S}, using a stacking technique for sources in the NVSS, found a lower value of $\leq2.5$\,\%. \citet{2015ApJ...806...83O} found a median fractional polarisation for an AGN sample using the NVSS of 6.2\,\%. \citet{2010MNRAS.402.2792S} found a linear increase in FP with lower flux densities. \citet{2021arXiv210702492B} showed that this trend is also visible in the analyses by \citet{2010ApJ...714.1689G}, \citet{2014MNRAS.440.3113H}, and \citet{2007ApJ...666..201T} and originates from the sensitivity limited completeness of the survey.
        
\begin{figure}[htb]
                \includegraphics{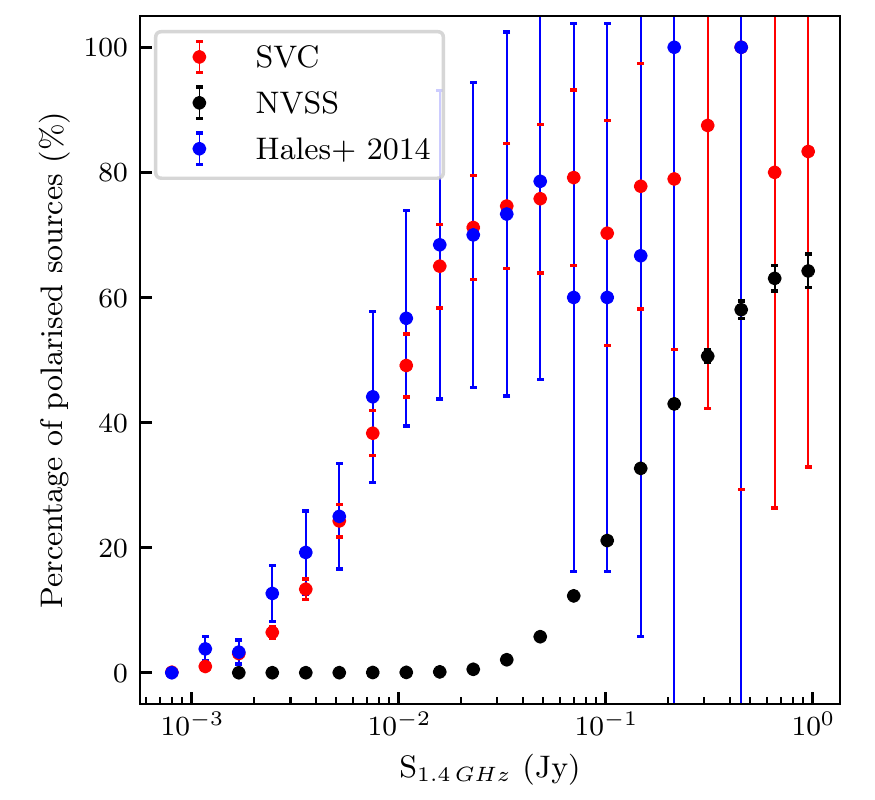}
                \caption{Total power emission $S_{1.4GHz}$ plotted against the percentage of sources showing polarised emission. The red, black, and blue dots show the values from this work, a combination of the two NVSS catalogues published by \citet{1998AJ....115.1693C} and \citet{2009ApJ...702.1230T}, and the results from \citet{2014MNRAS.440.3113H}.}
                \label{plot_TP_FSP}
        \end{figure}

        We compared the percentage of polarised sources with their flux in TP ($S_{1.4GHz}$) for our SVC data, the NVSS total and polarised source catalogues published in \citet{1998AJ....115.1693C} and \citet{2009ApJ...702.1230T}, respectively, and the catalogue of \citet{2014MNRAS.440.3113H} by counting the number of total power and polarised source components within a certain TP-flux bin (see Fig. \ref{plot_TP_FSP}). For comparability, the same bins were chosen for all catalogues and the Poisson error determined for each individual bin and catalogue. FP uncertainties were then calculated using Gaussian error propagation.
        
        All three catalogues show the same behaviour of a decreasing number of polarised sources detected with TP flux. The noise in PI images for radio observations is usually lower than for their accompanying TP images by a factor of $\sqrt{2}$  because of the combination of the two independent Stokes parameters Q and U. For a given source population with FP values of only a few per cent, the PI counterparts of TP detections are often hidden below the noise limits. This results in a large fraction of undetected polarised sources especially towards lower TP fluxes, if we assume a constant FP over the whole total power flux range.
        
        The maximum fraction of polarised sources  is $\approx80\,\%$ for our data and those of \citet{2014MNRAS.440.3113H}, while the NVSS data show a maximum of $\approx60\,\%$. As expected, due to the lower sensitivity of the NVSS, the decrease in percentage of polarised sources detected already starts at fluxes of $S_{1.4GHz}\approx0.8$\,Jy and reaches nearly 0\,\% at $S_{1.4GHz}\approx10^{-2}$\,Jy. Comparing our data with those of \citet{2014MNRAS.440.3113H} we notice a decrease starting at $S_{1.4GHz}\approx2\times10^{-2}$\,Jy, which reaches an absolute minimum value of 0\,\% at $S_{1.4GHz}\leq10^{-3}$\,Jy. A combination of effects can be responsible for these differences between the NVSS and the SVC-/\citet{2014MNRAS.440.3113H}-data, such as the different spatial and frequency resolutions of the surveys, the use of the RM-synthesis technique, or physical depolarisation effects. For a detailed discussion on this topic, we refer the reader to Berger et al. in prep.
        
        We notice a difference between our subsamples of resolved sources and unresolved sources with median fractional polarisations of $5.47\pm0.24$\,\% and $4.33\pm0.17$\,\%, respectively. This difference was previously found by \citet{2010ApJ...714.1689G}, who also measured a higher median fractional polarisation for resolved sources ($6.8\pm0.7$\,\%) compared to compact ones ($4.4\pm1.1$\,\%).
        
        \begin{table}[htb]
                \centering
                \caption{Cross-match statistics for the SVC data.} 
                \label{table:SVC_crossmatches}
                \begin{tabular}{ccccc}
                        \toprule
                        \toprule
                        Field & N$_{NVSS}$ & N$_{WISE}$ & N$_{SDSS}$ & N$_{SDSS,z}$ \\
                        \midrule
                        SVC1 & 4 & 146 & 54 & 5 \\
                        SVC2 & 4 & 207 & 121 & 26 \\
                        SVC3 & 15 & 210 & 42 & 10 \\
                        SVC4 & 16 & 207 & 1 & 0 \\
                        SVC5 & 13 & 197 & 101 & 21 \\
                        All & 52 & 967 & 319 & 62 \\
                        \bottomrule
                \end{tabular}
                \tablefoot{N$_{NVSS}$, N$_{WISE}$, and N$_{SDSS}$ are the number of cross-matches found in the NVSS-, AllWise-, and SDSS catalogues. N$_{SDSS,z}$ is the number of cross-matched SDSS sources where a redshift is given.}
        \end{table}
        
        We found 967 counterparts for our polarised sources in the AllWise catalogue and 319 counterparts in the SDSS DR16 release, of which 62 have spectroscopic redshift estimates. We note that 52 of our sources were previously known from the NVSS RM catalogue. See Table \ref{table:SVC_source_statistics} for an overview of the general statistics of the SVC data, and see Table \ref{table:SVC_crossmatches} for details of the cross-matches. 
        
        \subsection{Comparison with NVSS}
        
        In the following, we compare our PI-, RM-, and FP values for the 52 sources for which we were able to identify NVSS counterparts with literature values from their data to further verify the reliability of our catalogues. Of these sources, 35  are classified as resolved in the SVC, which is a fraction of 67.3\%. This is higher  than the average fraction of resolved sources of the whole SVC data set by a factor of approximately two. This is not surprising because the NVSS data are strongly biased towards bright sources, which have a higher probability of being closer and therefore appear resolved. The median ratio of PI between our detections and the NVSS counterparts is 1.18. If we only look at the unresolved sources, this drops to a nearly matching ratio of 1.03 while it rises for extended sources to 1.20. We explain this difference by the effect of beam depolarisation. Polarised sources in our survey can be resolved and components separated, which, in the NVSS with its approximately two times larger synthesised beam, become partly depolarised due to their differing magnetic field directions within the same source and resolution element.
        
\begin{figure}[htb]
                \includegraphics{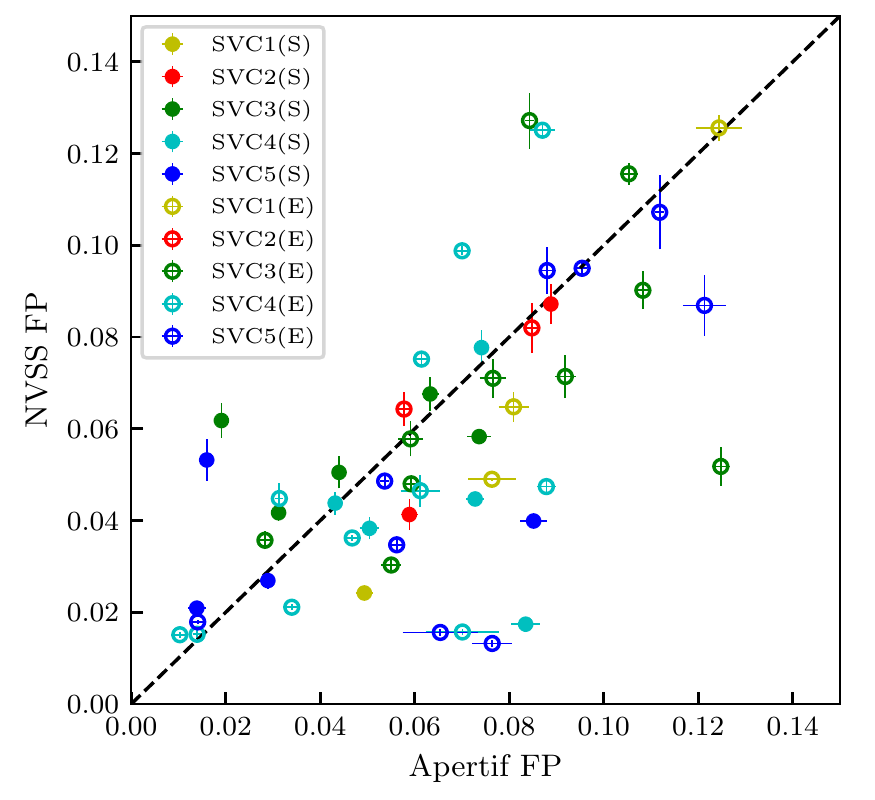}
                \caption{Fractional polarisation of the SVC sources plotted against the values of their NVSS counterparts from the NVSS RM catalogue \citep{2009ApJ...702.1230T}. Data from the different fields is marked by colour. Filled circles are unresolved sources while open circles represent resolved ones. The dashed black line represents the 1:1 ratio of the FP values.}
                \label{plot_FP_SVC_NVSS}
        \end{figure}

        For the FP, this behaviour is less pronounced  with a ratio of 1.06, but we can also trace a higher FP ratio between resolved and unresolved sources with values of 1.08 and 1.02, respectively. For the PI and FP ratios, we checked for any specific bias towards a certain SVC observation (see Fig. \ref{plot_FP_SVC_NVSS}); none could be found, and so we can again conclude that we see constant characteristics of the Apertif instrument for all SVC observations.
        
        We determined the median RM values and their standard deviation for each individual SVC field for all sources of a single field ($\widetilde{RM}$), only the unresolved sources ($\widetilde{RM}_S$), and only the resolved sources ($\widetilde{RM}_E$), as well as all sources with NVSS counterparts ($\widetilde{RM}_{NVSS}$). In addition, we cross-checked the central pointing positions from Table \ref{table:SVC_params} in the map of \citet{2012A&A...542A..93O}, which represents the RM caused by the Milky Way foreground magnetic field, and extracted the RM value ($RM_{MW}$) at this position. The results are compiled in Table \ref{table:SVC_RM_statistics}.
        
        \begin{table*}[htb]
                \centering
                \caption{Galactic coordinates and RMs for the SVC fields.} 
                \label{table:SVC_RM_statistics}
                \begin{tabular}{cccccccc}
                        \toprule
                        \toprule
                        Field & $l$ & $b$ & $\widetilde{RM}$ & $\widetilde{RM}_S$ & $\widetilde{RM}_E$ & $\widetilde{RM}_{NVSS}$ & $RM_{MW}$ \\
                        & [deg] & [deg] & [rad/m$^2$] & [rad/m$^2$] & [rad/m$^2$] & [rad/m$^2$] & [rad/m$^2$] \\
                        \midrule
                        SVC1 & 95.681 & -22.349 & $-126.19\pm85.31$ & $-103.81\pm87.77$ & $-155.42\pm54.08$ & $-123.29\pm93.68$ & $-131.07\pm25.95$ \\
                        SVC2 & 100.939 & 60.546 & $9.71\pm33.68$ & $15.41\pm35.36$ & $8.41\pm30.19$ & $16.94\pm11.79$ & $11.82\pm4.81$ \\
                        SVC3 & 137.785 & -27.078 & $-46.59\pm38.21$ & $-41.46\pm40.30$ & $-48.34\pm33.11$ & $-55.40\pm29.75$ & $-47.87\pm14.88$ \\
                        SVC4 & 97.789 & -17.882 & $-159.78\pm104.27$ & $-156.20\pm108.58$ & $-168.15\pm96.96$ & $-127.87\pm99.84$ & $-220.52\pm22.77$ \\
                        SVC5 & 65.517 & 69.869 & $0.28\pm30.02$ & $0.55\pm29.72$ & $-0.73\pm30.88$ & $0.75\pm25.12$ & $2.84\pm3.54$ \\
                        \bottomrule
                \end{tabular}
                \tablefoot{$l$ and $b$ are the Galactic longitude and latitude coordinates for each SVC field calculated from the RA-/DEC-coordinates given in Table \ref{table:SVC_params}, respectively. $\widetilde{RM}$, $\widetilde{RM}_S$, and $\widetilde{RM}_E$ are the median rotation measures and their standard derivation of all polarised sources, unresolved polarised sources and resolved polarised sources of a single SVC field, respectively. $\widetilde{RM}_{NVSS}$ is the median rotation measure and its standard deviation of all sources where an NVSS counterpart could be identified. $\widetilde{RM}_{MW}$ is the rotation measure and its error extracted from the map of \citet{2012A&A...542A..93O} for the given $l$- and $b$-coordinates.}
        \end{table*}
        
\begin{figure}[htb]
                \includegraphics{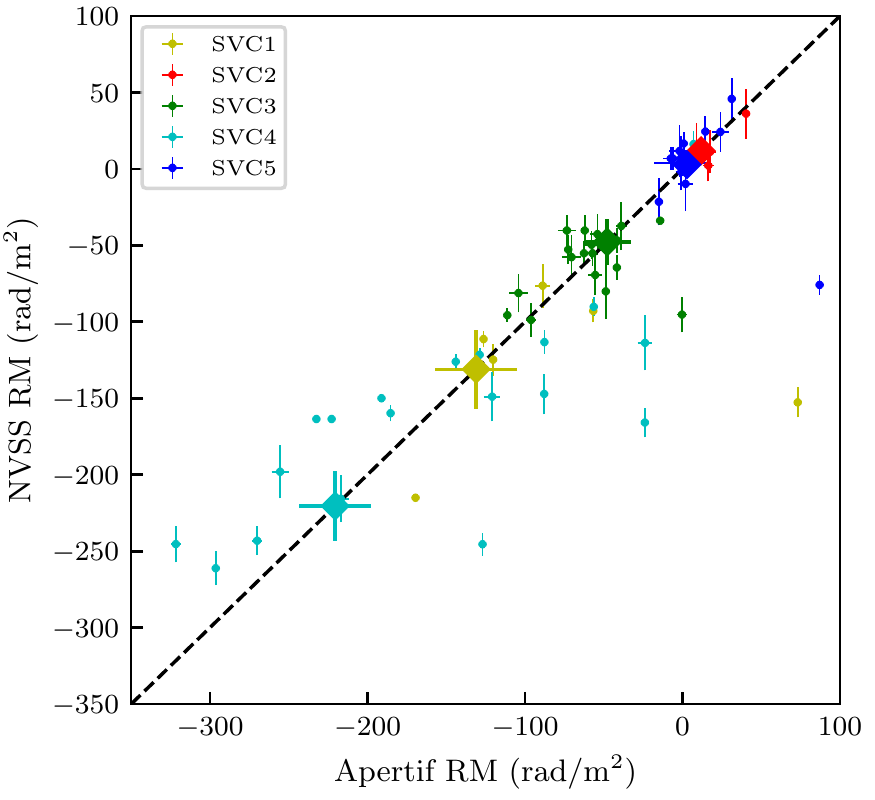}
                \caption{Rotation measures of the SVC sources plotted against the values of their NVSS counterparts from the NVSS RM catalogue \citep{2009ApJ...702.1230T}. Data from the different fields are marked with different colours. The diamonds show the RM value extracted from \citet{2012A&A...542A..93O} for the central pointing positions (see Table \ref{table:SVC_params}) of the five SVC fields. The dashed black line represents the 1:1 ratio of the RM values.}
                \label{plot_RM_SVC_NVSS}
        \end{figure}

        Except for field SVC4, we can see an agreement within a range of 10\,rad/m$^2$ between $\widetilde{RM}_{NVSS}$ and $RM_{MW}$, confirming the reliability of our measurements compared with the reference values in the NVSS RM catalogue. Figure \ref{plot_RM_SVC_NVSS} shows a direct comparison between the individual RM values for the SVC detections and their NVSS counterparts. We see that sources for individual fields are clustering around certain RM values. This is not surprising as the RM of most of our sources mainly originates from the magnetic field in the Milky Way foreground and is not intrinsic to the sources themselves, meaning that regions of several square degrees show very similar values. Sources in field SVC4 show a larger scatter compared to their NVSS RM values. We want to mention two possible reasons for this. Firstly, the field is the closest one in projection to the Milky Way disc out of the five SVC fields, where the magnetic field of the Milky Way foreground becomes more turbulent and therefore RM values on square-degree scales show greater variation. Secondly, the NVSS observations were taken with two  bands of 50\,MHz in width with only a single channel, meaning that the sensitivity of these observations might be affected by bandwidth depolarisation. This effect would lead to smaller PI values for sources with higher RMs.
        
        \subsection{Usability for RM gridding}
        
        Rotation-measure grids are one of the main tools for tracing the Milky Way \citep{2012A&A...542A..93O,2020A&A...633A.150H} and intergalactic magnetic fields \citep{2004NewAR..48.1003G}. These fields are usually too weak for a direct detection. Therefore, polarised background sources are used, where the intervening magnetic field is adding its RM component to the one of the background source. The precision of the RM grid is given by the source density and (S/N) of the individual polarised source detections as well as their uncertainties.
        
        Comparing $\widetilde{RM}$ and $RM_{MW}$ we see an agreement of less than 5\,rad/m$^2$. This shows that most of the sources in our catalogue are suitable for tracing the Milky Way magnetic field. Their sheer number, even though the sources are generally fainter, can help to more precisely determine the RM of the Milky Way foreground. The median RM values for unresolved and resolved sources of individual fields differ by up to $\approx12$\,rad/m$^2$ for fields SVC2 to SVC4. For SVC1, this value differs by $\approx52$\,rad/m$^2$. The RMs of extended sources are often partly influenced by their intrinsic magnetic fields. We want to note that for field SVC1, but also for fields SVC2 and SVC3, $RM_{MW}$ lies in between $\widetilde{RM}_S$ and $\widetilde{RM}_E$. For further discussion on this, we refer to Sect. \ref{Sect_discussion}.
        
\begin{figure}[htb]
                \includegraphics{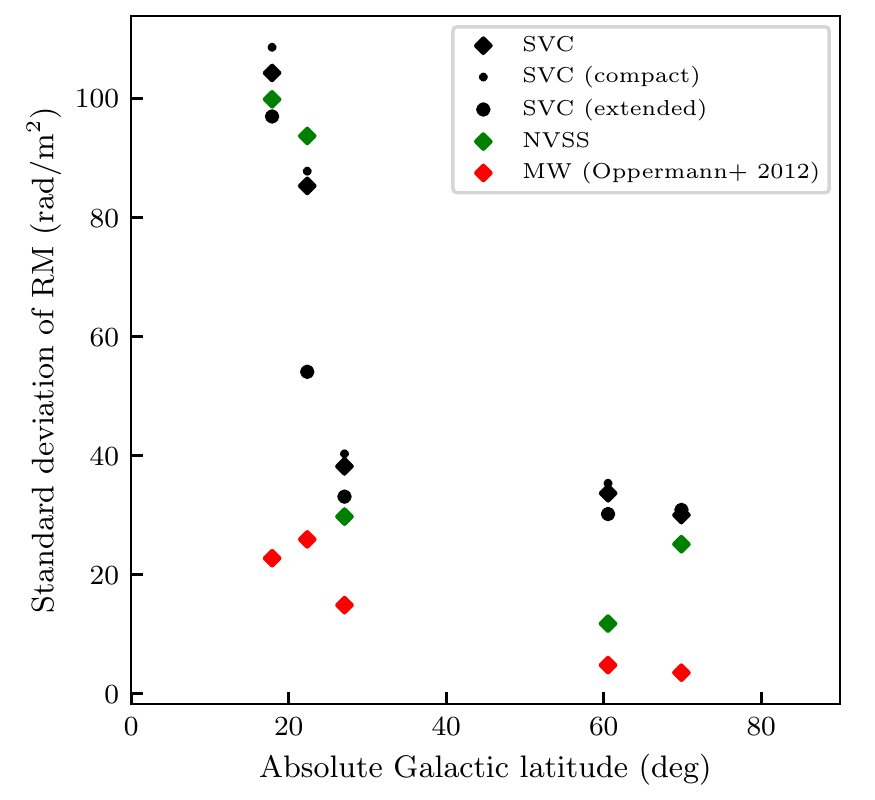}
                \caption{Absolute Galactic latitude for the five SVC fields plotted vs. the standard deviation of the RMs of the sources in the field. Black diamonds show the values for compact and extended sources in an SVC field, small circles show only compact sources, and large circles show only extended sources. Green diamonds show only the values for the cross-matched NVSS sources and the red diamonds the Milky Way foreground RM from \cite{2012A&A...542A..93O}.}
                \label{plot_RM_MW}
        \end{figure}

        The five SVC fields cover absolute Galactic latitudes of $17.8<|b|<69.9$. We observe higher absolute median RMs for sources closer to the Galactic plane as well as higher standard deviations  by a factor of up to three (see Fig. \ref{plot_RM_MW}). Regions closer to the Milky Way plane usually show higher absolute median RMs. The stronger magnetic field and higher thermal particle density causes a higher rotation of the electric field vector on the line of sight. In addition, these fields are usually more turbulent and a higher variation of the RMs of sources within these fields is expected.
        
        With our measurements of $RM_{MW}$ we can confirm the above expectations. Therefore, we state that the variation within our fields is dominated by the physical variations of the magnetic field of the Milky Way itself and is not driven by uncertainties in our measurements. 
        
        \section{Source characteristics}
        \label{sect_source_characteristics}
        
        In the following, we aim to investigate the types of sources in our source sample. The additional IR and optical information in our catalogues not only allows us to distinguish our sources by their relative brightness or morphology, but also by their host galaxy type, star-formation properties, and absolute radio brightness. This allows us a statistical analysis of the types of sources and the characteristics of the host galaxies, which dominate the polarised sky down to $\upmu$Jy-levels.
        
        \subsection{AGN and star-forming galaxies}
        \label{subsect_AGN_SF}
        
        Polarised emission in galaxies can have two different origins, namely the radio lobes generated by SMBHs situated in the centres of galaxies or the superbubbles and/or outflows created by star formation. For the latter type, polarised emission has only been detected in a handful of nearby galaxies \citep{2015A&ARv..24....4B} while all polarised sources further away have so far been classified as AGN. In the following, we aim to investigate our sample in view of a possible contribution from star formation.
        
        IR colours are known to be a good tracer of ongoing star formation \citep{2009ApJ...692..556R} because of their good sensitivity to heated dust particles in star-forming regions. One of the famous connections between star formation and FIR luminosity is the several orders of magnitude in luminosity spanning FIR--radio correlation \citep{1985A&A...147L...6D}. While this close correlation was first only shown to hold for nearby galaxies, later studies showed that it does not evolve up to redshifts of $z\approx4$ \citep{2015ApJ...807..141P}. \citet{2016ApJ...818..182V} found that the correlation can be extended towards the mid-infrared (MIR) WISE bands for galaxies with redshifts up to $z\approx0.03$. \citet{2009ApJ...692..556R} showed that 24\,$\upmu$m-fluxes can be used to estimate star-formation rates (SFRs) to within an order of magnitude for redshifts of up to $z\approx2$.  \citet{2017ApJ...850...68C} estimated that the WISE 12$\upmu$m- and 22$\upmu$m-bands are a good tracer of star formation over nearly five orders of magnitude. We now want to investigate our sample for possible objects dominated by star formation. Here we assume that the MIR--radio correlation holds as well as  the FIR--radio correlation for similar z ranges.
      
        \begin{figure*}[htb]
                \includegraphics{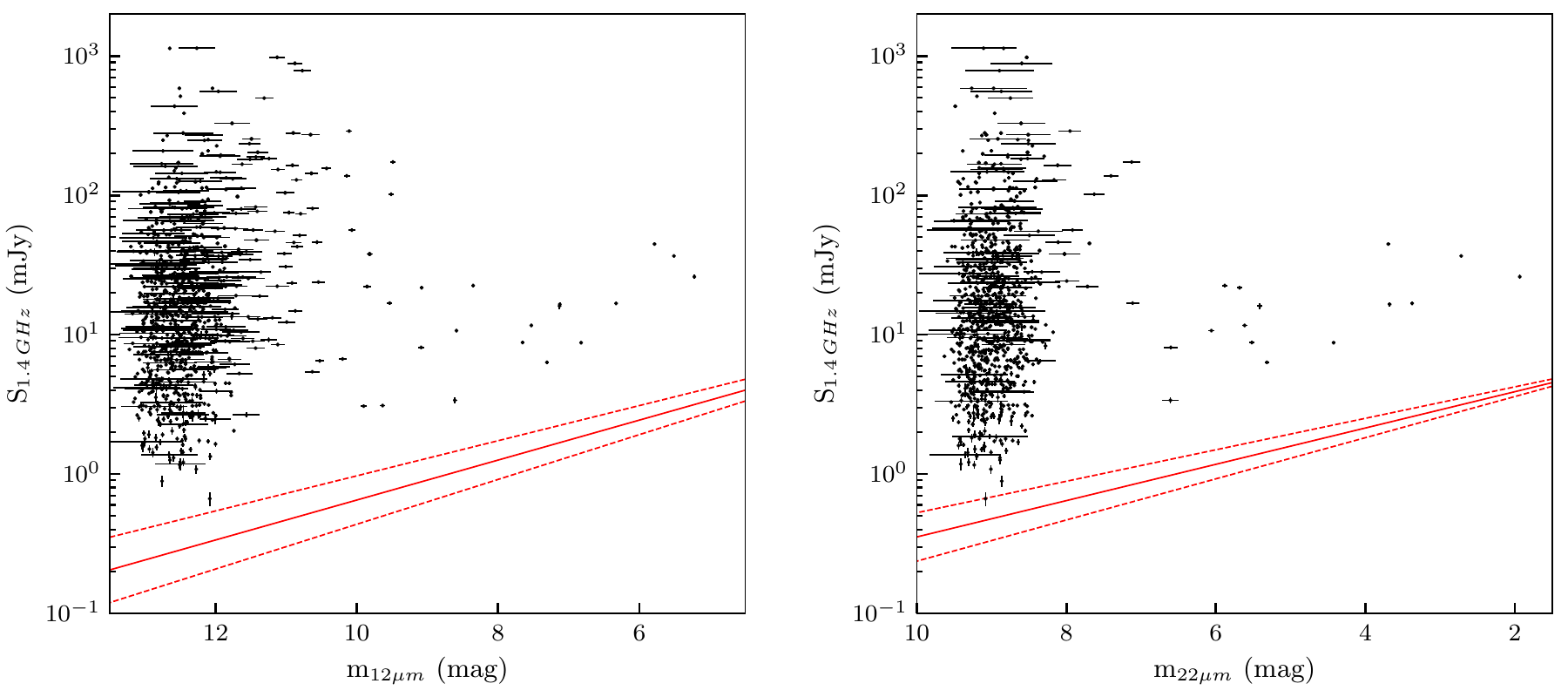}
                \caption{AllWISE 12$\upmu$m (left) and 22$\upmu$m (right) brightness vs. our radio luminosity for all polarised sources with AllWISE counterparts. AllWISE values are given in Vega magnitudes. The solid red line represents the MIR--radio correlation from \citet{2016ApJ...818..182V} with the dotted lines showing its uncertainties.}
                \label{plot_radio_MIR_correlation}
        \end{figure*}
        
        Figure \ref{plot_radio_MIR_correlation} shows the MIR brightness in the AllWISE 12$\upmu$m and 22$\upmu$m bands plotted against their total radio luminosity. Only one source (APSVC\_342.354+38.810 in SVC4) lies within the errors, which after closer inspection shows a very faint MIR counterpart and a high FP of $\sim$26\,\%. For a detection of polarised emission with a star-forming origin, we would expect very low values of FP. We therefore conclude that our sample is dominated by AGN-driven radio emission.
        
        To investigate the MIR brightness contributed by star formation in the host galaxies of our sample, we want to use criteria that are independent of the radio measurements. Several authors have been able to successfully distinguish between AGN- and star-formation-dominated sources using only the WISE MIR colours \citep{2011ApJ...735..112J,2012ApJ...753...30S,2012MNRAS.426.3271M}. All of these latter authors used empirical criteria for the colour relations between the WISE 3.4$\upmu$m-, 4.6$\upmu$m-, and 12$\upmu$m brightnesses. While \citet{2012ApJ...753...30S} used a simple limit of $m_{3.4\mu m}$ - $m_{4.6\mu m} > 0.8$, \citet{2011ApJ...735..112J} and \citet{2012MNRAS.426.3271M} defined limits that have a quadrangle- and wedge-shape, respectively. \citet{2012MNRAS.426.3271M} showed that the additional information of the WISE 22$\upmu$m-band does not improve the quality of the selection criteria, mostly because of the usually lower S/N in this band compared to the other three.
        
        In order to distinguish between sources with high S/N (HS/N) and sources with low S/N (LS/N), we set a S/N limit of $>5\sigma$ in all used WISE bands to classify sources as HS/N. This criterion was also used by \citet{2012MNRAS.426.3271M} for their investigation. Of our 967 sources, 101  fall into the HS/N category while 866 are identified as having LS/N. One, 142, and 865 sources are categorised as LS/N because of the 5$\sigma$ limit in the WISE 3.4$\upmu$m, 4.6$\upmu$m, and 12$\upmu$m bands, respectively. Therefore, the high number of LS/N sources is mostly caused by the lower sensitivity in the WISE 12$\upmu$m band.
        
        \begin{figure}[htb]
                \includegraphics{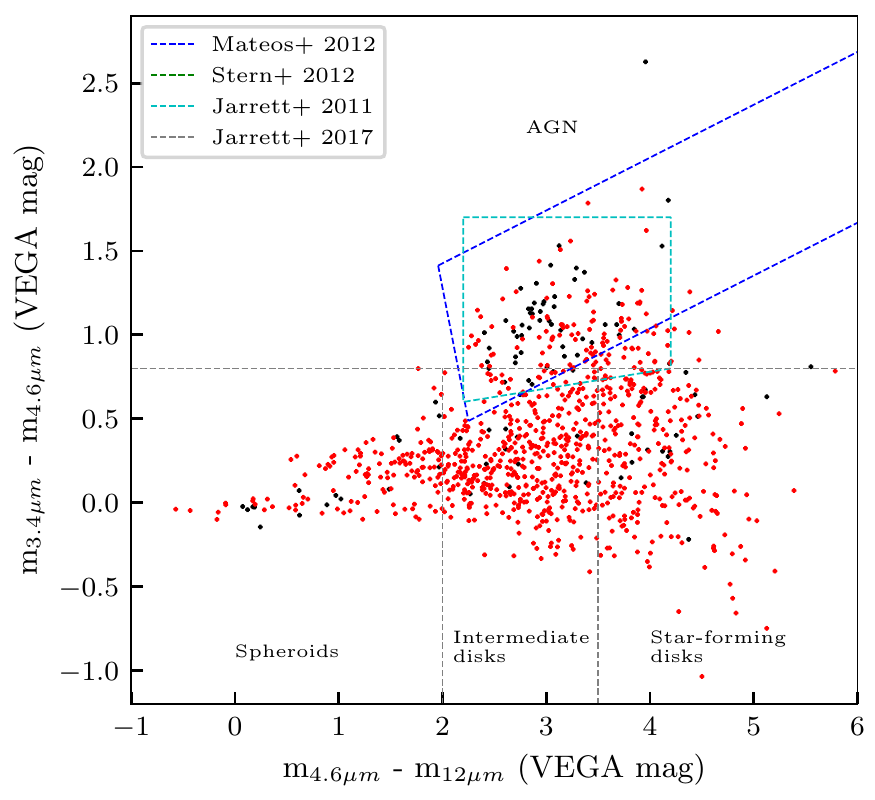}
                \caption{AllWISE 3.4$\upmu$m - 4.6$\upmu$m brightness vs. 4.6$\upmu$m - 12$\upmu$m brightness. All values are given in Vega magnitudes. Black and red dots represent HS/N and LS/N sources, respectively. The green, cyan, and blue dashed lines illustrate the AGN-selection criteria from \citet{2012ApJ...753...30S}, \citet{2011ApJ...735..112J}, and \citet{2012MNRAS.426.3271M}, respectively. The grey dashed lines divide the diagram into regions where AGN, spheroids, intermediate disc galaxies, and star-forming galaxies are positioned following the criteria in \citet{2017ApJ...836..182J}.}
                \label{plot_AGN_selection_MIR}
        \end{figure}
        
        Figure \ref{plot_AGN_selection_MIR} shows the relation between $m_{3.4\mu m}$ - $m_{4.6\mu m}$ and $m_{4.6\mu m}$ - $m_{12\mu m}$ for all polarised sources with IR counterparts. The overall distribution of our sample looks very similar to that of the sample in \citet{2012MNRAS.426.3271M}. The selection criteria from \citet{2012ApJ...753...30S} (green line), \citet{2011ApJ...735..112J} (cyan quadrangle), and \citet{2012MNRAS.426.3271M} (blue wedge) show very similar results. Of the 101 sources with a HS/N, 49, 51, and 52  lie within the AGN-selection criteria defined by \citet{2012ApJ...753...30S}, \citet{2012MNRAS.426.3271M}, and \citet{2011ApJ...735..112J}, respectively, while only 115, 103, and 133 out of 866 sources are situated inside the AGN regions for the LS/N sample, respectively. Most sources not fulfilling these criteria are located in the range of $-0.5\leq m_{3.4\mu m} - m_{4.6\mu m} \leq 0.5$. Using templates from \citet{2008ApJ...675..960P} for ultra-luminous infra-red galaxies (ULIRGs) and starburst sources and a template from \citet{2002ApJ...576..159D} for star-forming spiral galaxies, \citet{2012MNRAS.426.3271M} showed that this region is foremost dominated by elliptical and star-forming galaxies. The large number of LSNR sources in this area is mostly caused by the uncertainty in $m_{12\mu m}$, which only affects the x-position of sources in Fig. \ref{plot_AGN_selection_MIR}. Therefore, this uncertainty cannot move sources from the lower area of Fig. \ref{plot_AGN_selection_MIR} up into the AGN area, meaning that a misclassification of elliptical- and star-forming-dominated hosts is very unlikely.
        
        \citet{2017ApJ...836..182J,2019ApJS..245...25J} and \citet{2017MNRAS.464.1306C} used the same colour--colour diagrams to separate hosts of radio galaxies according to their morphological type. In the following, we use the criteria from \citet{2017ApJ...836..182J} to distinguish between AGN, spheroids, intermediate discs, and star-forming discs. A WISE colour of $m_{4.6\mu m}$ - $m_{12\mu m} < 2.0$ indicates spheroids, $2.0 \leq m_{4.6\mu m}$ - $m_{12\mu m} < 3.5$ indicates intermediate disc galaxies, and $m_{4.6\mu m}$ - $m_{12\mu m} \geq 3.5$ indicates star-forming disc galaxies. Hosts with $m_{3.4\mu m}$ - $m_{4.6\mu m} > 0.8$ are classified as AGN. The four different regions are shown in Fig. \ref{plot_AGN_selection_MIR} as grey dashed lines. We find 164 AGNs (17.0\,\%), 143 spheroids (14.8\,\%), 421 intermediate (43.5\,\%), and 239 star-forming disc galaxies (24.7\,\%). Comparing our numbers to the results by \citet{2015ApJ...806...83O} ---who used the NVSS database and therefore mostly bright polarised sources to determine host galaxy types from polarisation data--- shows a significant difference. While our sample is dominated by intermediate and star-forming disc galaxies, their sample consists mainly of spheroidal galaxies. While sources in the faint sky are more likely to be located at higher redshifts, \citet{2017ApJ...836..182J} showed that for a redshift of $z=0.5,$ data points are moved towards the upper left in Fig. \ref{plot_AGN_selection_MIR} by values of $m_{3.4\mu m}$ - $m_{4.6\mu m}=0.1$ and $m_{4.6\mu m}$ - $m_{12\mu m}=0.5$. Therefore, even a large contribution of higher redshift objects cannot explain this discrepancy.
        
        \subsection{Redshift determination and distribution}
        \label{subsect_redshift}
        
        In order to further investigate the characteristics of our polarised sources, we need estimates of their distances. Only 62 out of our 1170 sources have direct spectroscopic redshift data from the SDSS database. In order to enlarge our sample, we fitted spectral energy distribution (SED) templates to the SDSS and allWISE values from our cross-matches using the Galaxy redshifts and physical parameters (GAZPAR\footnote{\url{https://gazpar.lam.fr/home}}) service. For this purpose, the Photometric Analysis for Redshift Estimate package (Le PHARE\footnote{\url{http://lephare.lam.fr}}) was used \citep{2006A&A...457..841I,1999MNRAS.310..540A} with the BC03-template \citep{2003MNRAS.344.1000B}, which is known to give a good representation of star-forming galaxies and AGN. The allWISE 12$\upmu$m and 22$\upmu$m data were not used because the BC03-templates do not cover these wavelengths. We were able to determine photometric redshifts for all our 319 sources where SDSS data were available. Of those, 257  did not have spectroscopic redshifts beforehand.
        
        \begin{figure}[htb]
                \includegraphics{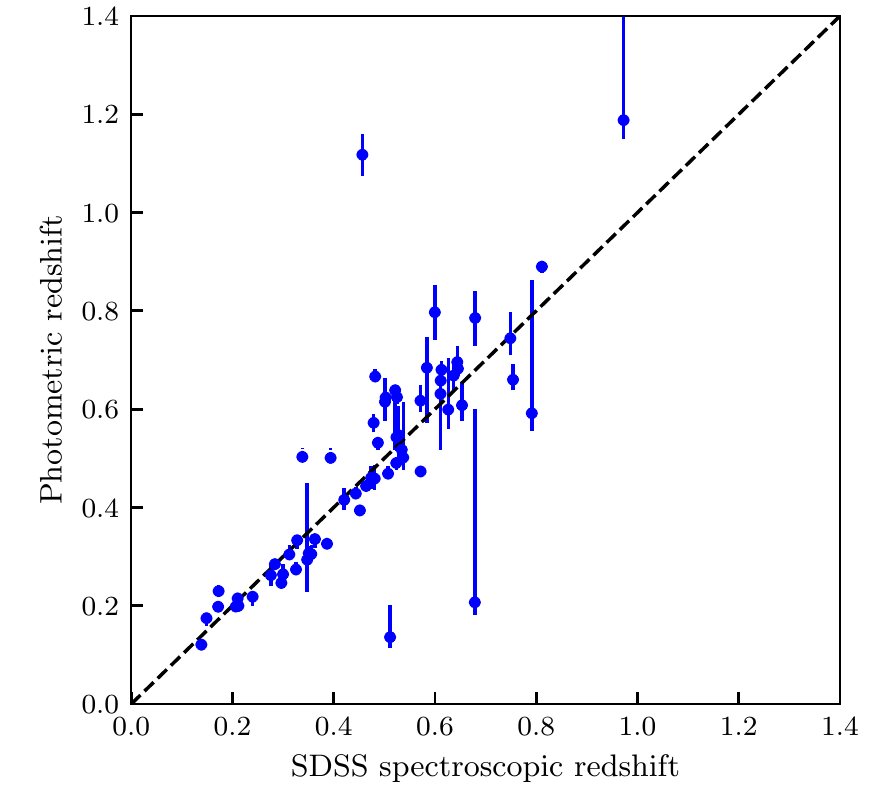}
                \caption{Spectroscopic redshift from the SDSS database vs. photometric redshifts from the SED-fitting. The dashed black line represents the 1:1 ratio between the two values.}
                \label{plot_specz_photoz}
        \end{figure}
        
        To verify the SED-fitting results, we compare the spectroscopic redshifts from the SDSS sample with our fitted values (see Fig. \ref{plot_specz_photoz}). For most of our sources, the two values deviate by no more than $\Delta z=0.1$, and so we conclude on the robustness of the photometric redshift determination.
        
        The redshift distribution of our sample is influenced by the completeness of the underlying SDSS information and the observational biases of this survey. To evaluate these parameters, we compare our data with the simulations of the Tiered Radio Extragalactic Continuum Simulation (T-RECS) \citep{2019MNRAS.482....2B}. We downloaded their AGN catalogue (\textsc{agnswide.dat}) which simulates the distribution of AGN sources over an area of 400 square degrees down to a flux limit of 100\,nJy at 1.4\,GHz. For comparability, we discarded all sources below our average detection limit of 75\,$\upmu$Jy in PI (see. Tbl. \ref{table:SVC_params} and Sect. \ref{subsection_source_finding}) from their catalogue. The average source density per redshift bin $\Delta z$ was then calculated, where we used $\Delta z=0.1$ and areas of 400\,deg$^2$ for the T-RECS dataset and 36.94\,deg$^2$ for the SDSS coverage of our data (see Sect. \ref{subsect_cross_matching}).
        
        \begin{figure}[htb]
                \includegraphics{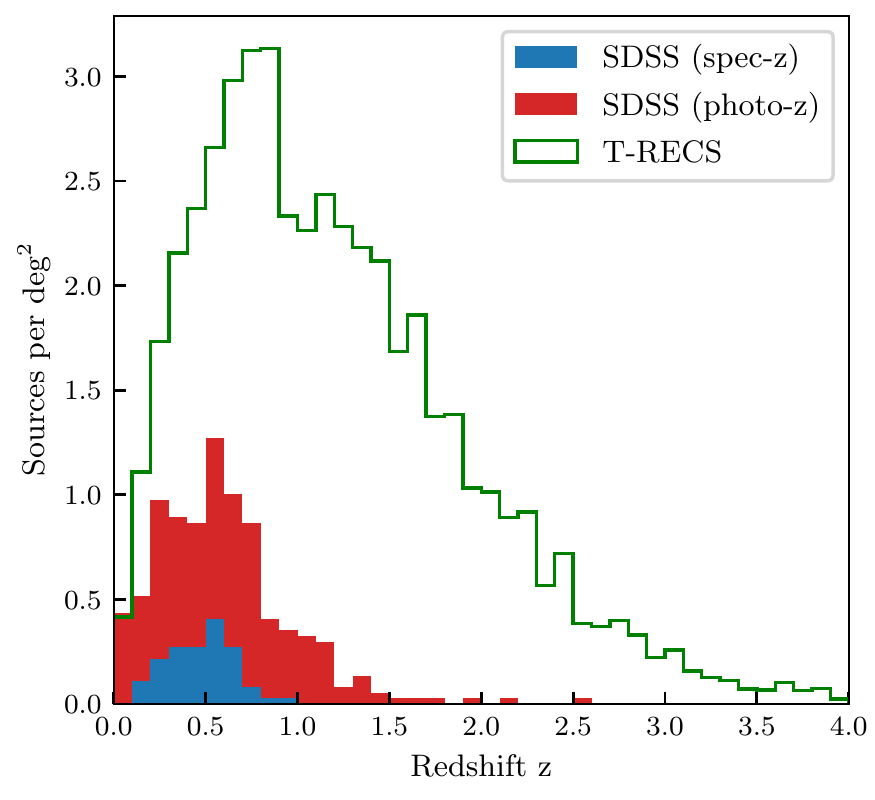}
                \caption{Redshift distribution vs. source density. Redshift bins are $\Delta z=0.1$. The blue filled histogram shows the distribution of sources with spectroscopic redshift information in the SVC-sample. The red histogram shows the distribution of sources with photometric redshift estimates from our SED fitting. These two histograms are stacked. The green line was derived from the wide AGN-catalogue in \citet{2019MNRAS.482....2B}.}
                \label{plot_SVC_TRECS_z}
        \end{figure}
        
        The results are shown in Fig. \ref{plot_SVC_TRECS_z}. Even though we are dealing with low absolute numbers of sources in our sample, the overall shape of the distribution for redshifts up to $z=0.6$ fits the simulations. It is remarkable that we find the exact source density predicted by the T-RECS simulations for redshifts up to $z=0.1$. We see a continuous decrease in the absolute number density of detected sources with respect to simulated ones with increasing redshift. While this discrepancy is approximately a factor of 2.5 for redshifts of $z=0.6$, it increases to a factor of six at redshifts of $z=1.2$. For higher redshifts, we only detect a very small number of sources. This mostly results from the fact that not all radio sources are equally bright in the optical, meaning that our subsample with optical counterparts is most likely biased towards radio sources with bright optical counterparts. In addition, we see a lack of sources beyond redshifts of $z\geq0.6$ even though the strong decrease beyond $z\geq0.8$ is still represented. There are far fewer sources with determined redshift at this cosmic distance for the SDSS sample as well, which is mostly due to sensitivity limitations \citep{2019MNRAS.484.1021O}.
        
        \subsection{Radio brightness and Fanaroff-Riley classification}
        
        AGN can generally be classified as radio-loud or radio-quiet, which also gives an indication of their activity. \citet{1989AJ.....98.1195K} defined the border between radio-loud and radio-quiet AGN by the ratio of the flux density at 5\,GHz $P_{5GHz}$ and the optical B-band $S_{B-band}$. Objects with ratios of $P_{5GHz}/S_{B-band} \geq 10$ are radio-loud and sources below this level are radio-quiet. While the former authors defined this criteria for Type 1 broad-line AGN, \citet{2010AJ....139.1089L} and \citet{2007ApJ...658..815S} extended these criteria to Type 2 narrow-line AGN, Seyfert galaxies, low-ionisation nuclear emission-line regions (LINERs), FRI radio galaxies, and optically selected quasars. Often radio-loud sources fall into the FRII category, while silent ones are more common for FRI sources. \citet{1994ASPC...54..319O} and \citet{1996AJ....112....9L} showed for a sample of AGN observed at 1.4\,GHz that the two FR phenotypes populate different areas in a diagram of optical versus radio brightness.
        
        In order to investigate the source distribution for our sample, we need to calculate the absolute radio brightness $P_{1.4GHz}$ given in units of W Hz$^{-1}$. For this we use the standard equation
        
        \begin{equation}
        P_{1.4GHz} = 4\pi D_L^2 S (1+z)^{-\alpha+1}
        ,\end{equation}
        
        where $D_L$ is the luminosity distance, $S$ is the total flux density at 1.4\,GHz in Jy, $z$ is redshift, and $\alpha$ is the spectral index. For our analysis, we limit our sample to sources where redshift information from SDSS is available. In addition, we assume a canonical value of $\alpha=-0.7$ from \citet{2002AJ....124..675C} for the spectral index $\alpha$ with the definition $S\propto\nu^{-\alpha}$, where $S$ is the radio flux in Jy, and $\nu$ the observed frequency. 
        
        The SDSS does not observe directly in the optical B-band, and so we have to calculate the B-magnitudes from the u- and g-band values. We used the conversion equation given in \citet{2005AJ....130..873J}:
        
        \begin{equation}
        B = g + 0.17 (u-g) + 0.11
        .\end{equation}
        
        \begin{figure}[htb]
                \includegraphics{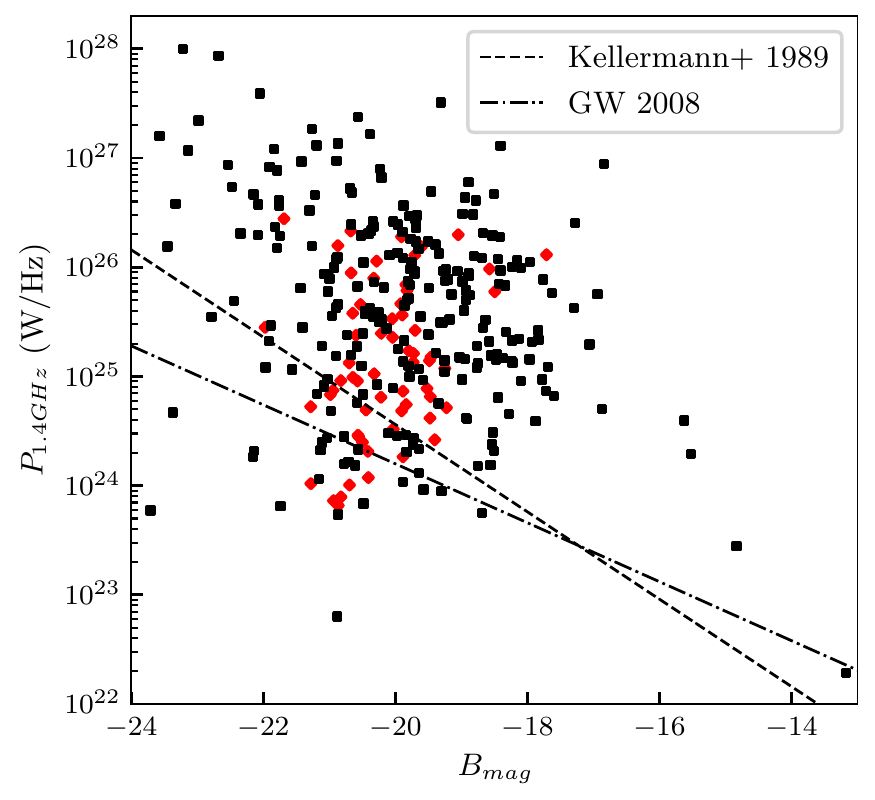}
                \caption{Absolute optical B-band magnitude $B_{mag}$  vs. absolute radio brightness at 1.4\,GHz $P_{1.4GHz}$. Red diamonds are sources with spectroscopic redshift information from the SDSS. For the black squares, photometric redshifts were determined (see Sect. \ref{subsect_redshift}). The border between radio-loud and radio-quiet AGN defined by \citet{1989AJ.....98.1195K} is plotted as a dashed line. The relation described in \citet{2008MNRAS.390..819G} to separate FRI and FRII objects is shown as the dash-dotted line.}
                \label{plot_AGN_luminosity}
        \end{figure}
        
        The final absolute optical brightness was then calculated using the distance modulus. The results are shown in Fig. \ref{plot_AGN_luminosity}. Of our 319 sources, 253 (79\%) and 66 (21\%)  are classified as radio-loud and radio-quiet, respectively, following the criterion of \citet{1989AJ.....98.1195K}. Only 40 (13\%) of our 319 sources fulfil the criterion for an FRI-classification while 279 (87\%) are classified as FRII \citep{1996AJ....112....9L}. We therefore conclude that our sample with optical counterparts and redshift information is dominated by radio-loud AGN mostly of the FRII type.
        
        \subsection{Influence of star formation}
        \label{subsect_influence}
        
        We show that, for a large part of our sample, the total radio flux is dominated by the activity of the AGN, but star formation can also play an important role. In order to estimate its contribution and its influence on the polarisation degree of the sources, we used the SED fitting described in Sect. \ref{subsect_redshift} to not only derive photometric redshifts, but also galaxy masses and SFRs of the host galaxies, independent of the radio measurements. While galaxy masses can be robustly derived by SED fitting, uncertainties for SFRs are higher and sometimes show unrealistically low values. Elliptical galaxies with the lowest known SFR are in the range of $10^{-4}$\,M$_\odot$yr$^{-1}$ \citep{2010MNRAS.402.2140S}. Therefore, we filtered out all objects with SFR values below $5\cdot10^{-4}$\,M$_\odot$yr$^{-1}$ and consider 198 out of 297 objects as valid.
        
\begin{figure}[htb]
                \includegraphics{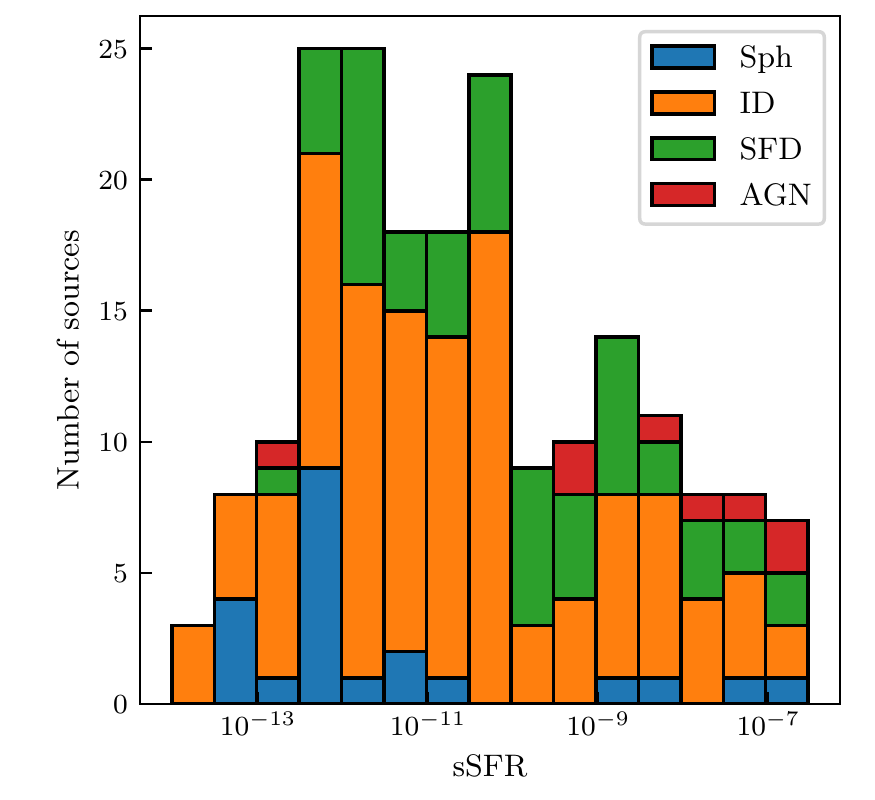}
                \caption{Histogram of the sSFRs estimated via SED fits to the host galaxies of the polarised sources distinguished between spherical galaxies (Sph), intermediate disc galaxies (ID), star-forming discs (SFD), and AGN using the criteria in Sect. \ref{subsect_AGN_SF}.}
                \label{plot_AGN_sSFR}
        \end{figure}

        Using the criteria described in Sect. \ref{subsect_AGN_SF}, we classified 22, 116, 52, and 8 sources as spheroidal galaxies, intermediate disc galaxies, star-forming disc galaxies, and AGN, respectively. To analyse the reliability of our SED fitting in view of SFRs and galaxy masses, we show the distribution of the specific star-formation rate (sSFR), that is the SFR per unit galaxy mass, in Fig. \ref{plot_AGN_sSFR}.
        
        As expected, spheroidal galaxies populate the lower end of the distribution mostly with sSFRs up to sSFRs $<10^{-11}$ \citep{2021arXiv210613812S} while AGN are mostly located at the high end with sSFRs $>10^{-9}$ \citep{2014MNRAS.443..755H}. Most of the star-forming and intermediate disc galaxies are found in the range of $10^{-12}<\textnormal{sSFR}<10^{-9}$ \citep{2007ApJ...658.1006M}.  
        
        To calculate the contribution of star formation to the absolute radio brightness $P_{1.4GHz}$, we use the calibration defined by \citet{2001ApJ...554..803Y}
        
        \begin{equation}
        \textnormal{SFR}_{1.4GHz} = 5.9 \times 10^{-22} P_{1.4GHz}
        ,\end{equation}
        
        where SFR is the total SFR in units of $M_\odot yr^{-1}$. We then define the radio excess parameter $\zeta$, which is the ratio between the SFR derived from SED-fitting ($\textnormal{SFR}_{fit}$) and the one calculated from the absolute radio brightness, as
        
        \begin{equation}
        \zeta = \frac{\textnormal{SFR}_{fit}}{\textnormal{SFR}_{1.4GHz}}
        .\end{equation}
        
        \begin{figure}[htb]
                \includegraphics{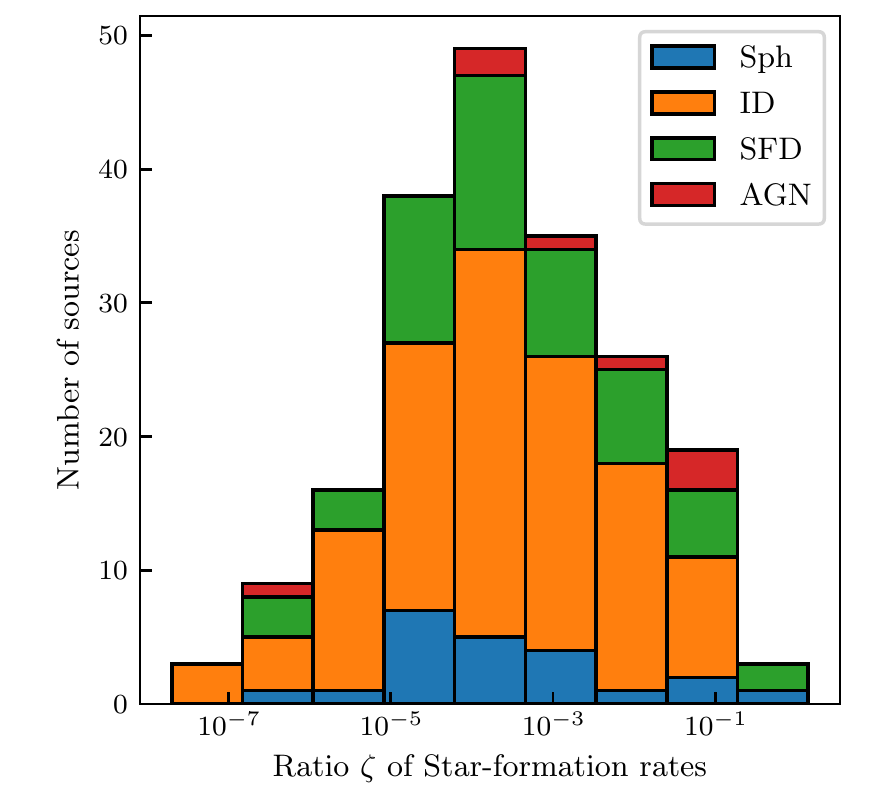}
                \caption{Histogram of the radio excess parameter $\zeta$, which denotes the ratios of the radio flux contribution of the star formation to the overall radio luminosity for our successfully SED-fitted sample of host galaxies.}
                \label{plot_SFRperc}
        \end{figure}
        
        Figure \ref{plot_SFRperc} shows that the $\zeta$-values span a range between $10^{-8} \leq \zeta \leq 1.3$. For most of our sources, $\zeta$-values are below $10^{-2}$. We can identify a peak at $\zeta$-values around $10^{-4}$. We find 29 sources for which the contribution lies above 1\,\% and four sources with $\zeta\geq0.1$. We therefore conclude that star formation is not contributing a significant part to the overall radio luminosity for approximately 90\% of our polarised sources. Surprisingly, we cannot identify a difference between the relative distributions of spheroidal, intermediate disc, star-forming disc, and AGN sources. 
        
        \begin{figure}[htb]
                \includegraphics{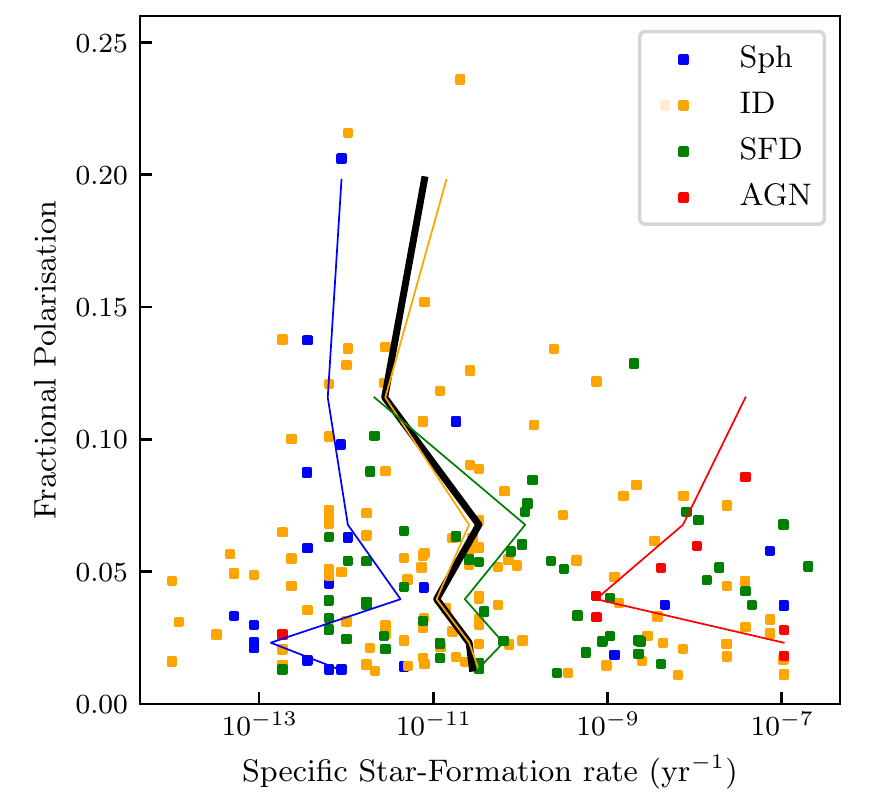}
                \caption{sSFR vs. the fractional polarisation of our host galaxies by galaxy type (see Sect. \ref{subsect_AGN_SF}). Data points are colour coded according to host galaxy type as described in Fig. \ref{plot_AGN_sSFR}. The solid lines represent the median values over logarithmically derived bins for each host galaxy type. The black line represents the median values for the whole galaxy sample.}
                \label{plot_SFR}
        \end{figure}
        
        We now want to further investigate the extent to which the star formation in the host galaxies of AGN amplifies or reduces the polarised emission detected in our sources. For this purpose, we compared the sSFR of our host galaxies to the measured FP values of the polarised radio emission (see Fig. \ref{plot_SFR}). The sSFRs span seven orders of magnitude between $10^{-7}$ and $10^{-14}$\,yr$^{-1}$. In order to overcome this large scatter, we binned the data over FP values between 0.1 and 0.25. This was performed on a logarithmic scale to mitigate a bias towards the much larger number of sources at small FPs compared to large ones. For each bin, the median was derived. 
        
        The medians for all our classified sources (black line in Fig. \ref{plot_SFR}) do not show any significant correlation or anti-correlation with sSFR or FP. Looking at individual types of sources following our classification scheme, we can clearly see a separation between the three groups of spheroidal galaxies, intermediate disc/star-forming disc galaxies, and AGN according to their sSFRs, as described earlier. Neither a correlation nor an anti-correlation can be observed between sSFR and FP for the individual types. For a detailed discussion, the reader is referred to Sect. \ref{Sect_discussion}.
        
        \subsection{Infrared faint radio sources}
        
        When comparing radio and IR fluxes, a specific class of objects show only very faint IR compared to their radio emission. Such sources were first detected by \citet{2006AJ....132.2409N} and are called IR-faint radio sources (IFRS). Later, \citet{2011ApJ...736...55N} found them to be radio-loud AGN with redshifts $z>1$. As our polarised source sample is most likely dominated by radio-loud AGN, and IFRS investigated by \citet{2011A&A...526A...8M} showed strong polarised emission \citep{2011A&A...526A...8M}, the contribution of IFRS to the faint polarised radio sky could be relevant. This applies especially for the fitting procedure described in Sect. \ref{subsect_influence}. IFRS sources are highly obscured, high-redshift galaxies. The SED of these sources most likely does not fulfill the typical criteria for the used fitting templates. A significant contribution of such sources to our sample could therefore corrupt the results.
        
        Criteria for classifying sources as IFRS were generalised by \citet{2011A&A...531A..14Z} in the following way: The flux ratio $S_{1.4GHz}/S_{3.6\mu m}$ exceeds a factor of 500 and the $S_{3.6\mu m}$ is lower than $30\,\upmu$Jy, where $S_{1.4GHz}$ and $S_{3.6\mu m}$ are the flux densities at 1.4\,GHz and at 3.6$\upmu$m in Janskys (Jy), respectively. To convert AllWISE magnitudes to Jy we used the relation given by \citet{2014MNRAS.439..545C}:
        
        \begin{equation}
        S_{3.4\mu m} = 306.682 \times 10^{(-m_{3.4\mu m/2.5})} \textnormal{Jy}
        .\end{equation}
        
        Of the 967 sources in our sample, 922  fulfill the cutoff criterion and 126 are then classified as IFRS by the flux density ratio. The number of resolved and unresolved sources is nearly equal with 58 and 68, respectively. \citet{2011A&A...531A..14Z} considered only unresolved radio sources without any signs of radio lobe structure as IFRS candidates. Later studies revealed that IFRSs are often high-z radio galaxies with double-lobed structures \citep{2017MNRAS.470.4956S}. In the following, we therefore analyse the resolved and unresolved sample of IFRS candidates independently. 
        
        Even though IFRSs are known to be very compact sources at high redshifts \citep{2019MNRAS.484.1021O}, \citet{2011A&A...526A...8M} showed that even at these distances, AGN sources with lobe structures of several kiloparsecs (kpc) can easily appear as resolved because of the nearly constant relation between apparent and linear size of 7\,kpc/arcsec for redshift values of $z>0.5$. For the average resolution of our observations, that is 20\,arcsec, and assuming that any source larger than one-fifth of the resolution would be classified as resolved, we can estimate the maximum physical size of unresolved sources to 28\,kpc. Using a combination of different radio surveys of two survey fields with a combined area of 1.8\,deg$^2$ and resolutions of between 4 and 9.4 arcsec, \citet{2017MNRAS.470.4956S} detected 14 unresolved IFRS sources and five resolved ones, which corresponds to an IFRS source density of 10.6 per deg$^2$. \citet{2011A&A...526A...8M} found three out of 17 IFRS sources to be polarised, which would correspond to a polarised IFRS source density of 1.86 per deg$^2$. Our polarised IFRS source density of  2.23 per deg$^2$ is a close match to this latter estimate.
        
        In addition to showing that IFRSs follow a similar redshift distribution to high-redshift radio galaxies, \citet{2019MNRAS.484.1021O} measured the redshift of 131 IFRSs. The z-range was found to lie between $z=1.63$ and $z=4.39$. The number of objects with measured spectroscopic redshifts determined by SDSS drops significantly beyond a value of $z>0.8$ \citep{2019MNRAS.484.1021O}. As a result, only one of our IFRS counterparts (APSVC\_211.418+54.187) has a spectroscopic redshift noted in the SDSS database with $z=0.46$. Closer inspection of the source reveals a double-lobe-dominated AGN with a diameter of 2 arcmin and a very faint IR counterpart in its centre, which corresponds to a physical size of the source of 707\,kpc. \citet{2017MNRAS.470.4956S} also reported sizes for their IFRSs of up to 330\,kpc, but at higher redshifts.
        
        If IFRS sources were to follow the same distribution as our whole sample, for which 62 SDSS redshifts are known, we would expect approximately seven sources with noted SDSS redshifts. As the high-z region is not covered by SDSS, a lower detection rate using the spectroscopic redshift data from SDSS for IFRS would confirm the high-z nature of these sources. On the contrary, we find 25 sources with photometric redshifts fulfilling the IFRS criteria. A very similar number of detections of approximately 28 sources is expected if the IFRS source distribution follows the same distribution as our complete photometric sample. We also compared the median redshift of both samples, finding $z=0.57\pm0.02$ for the whole sample and $z=0.60\pm0.12$ for the IFRS sample.
        
        Three IFRS sources from \citet{2011A&A...526A...8M} show polarised emission with FPs between 7\% and 12\%. For our IFRS sample, we report a median FP of $4.82\pm0.34$\% and do not see a significant difference between the polarised IFRS and polarised non-IFRS median FP values ($4.77\pm0.17$\%). We see the same difference between the values of unresolved and resolved IFRS sources as we noted for the whole sample (see Sect. \ref{subsect_source_statistics}) of $4.04\pm0.38$\% and $5.41\pm0.56$\%, respectively.
        
        Using the classification criteria above, a significant number of polarised sources ($\sim13\,\%$) in the faint polarised sky could fall into the IFRS category. Surprisingly, their polarised source characteristics do not seem to differ from the overall population of the faint polarised sky. This means that their contribution does not influence the results of our statistical analysis.
        
        \section{Discussion}
        \label{Sect_discussion}
        
        The low polarisation leakage  for the Apertif system and similarity of PI-, FP-, and RM values compared to reference sources using the NVSS database and Milky Way foreground maps allows us to reliably analyse polarisation data. Even though we excluded sources with FP values below the estimated leakage of 1\,\%, sources with overall low FP values are still more affected by leakage than sources with higher values and can introduce a bias towards the faint end of the polarised source population. In addition, the borders of our mosaics are imaged up to the 5\,\% primary beam level where leakages can rise to several percent (compare Fig. \ref{plot_leakage_response}). As only $\sim10\,\%$ of our overall imaged area lie in the 1\,\% leakage area and sources are less likely to be detected there due to the higher noise, we can exclude a major statistical contribution to our analysis. This is also supported by the similarity of our median FP values and source density compared to the publications by \citet{2014MNRAS.440.3113H,2010ApJ...714.1689G,2007ApJ...666..201T,2010MNRAS.402.2792S}. However, data of individual sources at the mosaic borders should be handled with care.    
        
        RMs of polarised background sources are often used for analysis of magnetic fields in nearby objects such as the Milky Way \citep{2012A&A...542A..93O,2020A&A...633A.150H} or the Magellanic Clouds \citep{2005Sci...307.1610G,2008ApJ...688.1029M}. The resolution of such studies and therefore the number and types of objects we are able to investigate using this RM gridding technique are strictly coupled to the density of polarised background sources. While the previous studies could only use a maximum of three polarised sources per square degree due to their limited sensitivity, we detect a polarised source density of 21 sources/deg$^2$. This source density is very similar to literature values found at the same frequency and with similar resolution. However, the exact value is always dependent on the primary beam response level up to which the data was imaged. To mitigate this effect, cumulative source counts should be used. Preliminary results show that the Apertif measurements (Berger et al. in prep.) are confirming the results shown in \citet{2014ApJ...785...45R}. The usability of a polarised source for an RM grid is also dependent on whether it is resolved or compact. \citet{2004NewAR..48.1289B} assumed that only 50\,\% of the detected polarised sources can be used. From our comparison of the standard deviations of the RM values of extended and compact sources, we can assume that, at least for mildly resolved sources, the RM value at their centres is a reliable probe of the intervening magnetic field and is not significantly influenced by their internal RM. Of our sources,  67\,\% show a compact morphology, and therefore the usable fraction of sources is likely to be larger than the value given by \citet{2004NewAR..48.1289B}.
        
        Literature values for the fraction of polarised sources are scattered between $\leq 2.5\,\%$ and 14.1\,\% (see \citet{2014MNRAS.440.3113H,2010ApJ...714.1689G,2007ApJ...666..201T,2014ApJ...787...99S,2010MNRAS.402.2792S,2015ApJ...806...83O}). \citet{2021arXiv210702492B} showed that the FP of sources is dependent on the completeness of the source sample as well as on its redshift distribution. In addition to this, we show that the FP of a source is dependent on whether it is compact or resolved with approximately 20\,\% more polarised emission originating from resolved sources.
        
        The independent combination of the two values in Stokes Q and U leads to a factor of $\sqrt{2}$ lower theoretical noise in the resultant PI images compared to the TP images. In addition, side-lobe confusion in TP images is usually higher due to the larger number of sources, which enhances the noise in those images. The use of individual channel images for generating PI mosaics ---where the primary beam response level is dependent on frequency--- adds additional noise at the borders and overlap regions of individual beams. Therefore, the fraction of polarised sources is often influenced by these factors, making a direct comparison between different instrumental setups difficult. As the FP and the fraction of sources for which polarised emission is detected are interconnected, the scatter seen in literature values is most likely caused by differences in the sensitivity and/or resolution of the individual surveys. 
        
        Differences in the FP values for compact and resolved polarised sources can have an instrumental and/or intrinsic physical origin. Polarised emission from sources with multiple components emitting at different polarisation angles leads to beam depolarisation. For our sources, which are mostly jet-dominated AGN, polarisation angles can change rapidly between the jets towards and within the lobes (see \citet{1988ARA&A..26...93S} and references therein for resolved polarisation studies of FRI- and FRII-sources). \citet{2015ApJ...806...83O} showed that sources with straight jets exhibit smaller FP values than sources with bent jets. \citet{2010ApJ...714.1689G} and \citet{1974A&A....34..341H} argued for a high percentage of lobe-dominated sources within their extended samples, which are known to exhibit higher degrees of polarisation.
        
        Physical origins are more complex. Internal and external Faraday dispersion can lead to a depolarisation of the signal by turbulent magnetic fields inside the sources themselves, in the environment surrounding them \citep{2014ApJS..212...15F}, or on the line of sight to the observer \citep{1966MNRAS.133...67B,1998MNRAS.299..189S}. These fields can be comprised of varying media, from turbulent cosmic magnetic fields over intracluster ones \citep{1988Natur.331..149L,1988Natur.331..147G} to magnetic fields within star-formation regions in the host galaxies.  
        
        The role of these star-forming regions in the polarisation properties of galaxies is not clear. While molecular clouds often host strong magnetic fields in the mG-regime \citep{2012ARA&A..50...29C} and star-forming galaxies can generate large-scale fields via star formation in combination with the $\alpha\Omega$-dynamo \citep{2015A&ARv..24....4B}, strong star formation, for example in starburst galaxies like M\,82, can lower the observed polarisation fraction due to a more turbulent magnetic field morphology \citep{2017A&A...608A..29A}. An interaction between AGN activity and molecular and neutral gas motions followed by star formation has been shown by \citet{2018A&ARv..26....4M}. The early phases of the radio jet propagation within the inner few kpc of the host galaxy environment are most important \citep{2021AN....342.1135M,2021A&A...656A..55M}. However, the influence of star formation on the polarised properties of the radio jets is under debate. \citet{2015ApJ...806...83O} investigated the differences in polarisation properties of radiative-mode AGN (high excitation radio galaxies (HERGs) and quasi-stellar objects (QSOs), which are known to host a radiatively efficient accretion disc and jet-mode AGN (low excitation radio galaxies; LERGs), which are inefficiently accreting. Even though HERGs are usually more powerful and larger objects, they showed lower median FPs than the smaller and less powerful LERGs. \citet{2015ApJ...806...83O} also argue that the differences in the polarisation properties of AGN are not intrinsic but rather influenced by the magnetic field configuration of their local environment.
        
        For our sample, we are not able to find a correlation or an anti-correlation of the FP with star-formation properties of the AGN host galaxies. This might be the consequence of a combination of the large range of sSFR we cover together with the rather small size of our sample. In addition, the derivation of SFR and sSFR from SED fitting using only a maximum of six photometric data points involves large uncertainties. Therefore, a larger source sample in combination with a finer selection of host galaxy types would shed light on this aspect.
        
        We find a significant number of sources classified as IFRS using the criteria by \citet{2011A&A...531A..14Z}. The definition of an IFRS is not consistent in the literature regarding the apparent size of these objects. While \citet{2019MNRAS.484.1021O,2011A&A...526A...8M} and \citet{2011ApJ...736...55N} defined them as being compact and mostly unresolved, \citet{2017MNRAS.470.4956S} also used the classification for resolved sources. The criterion by \citet{2011A&A...531A..14Z} was also used by the authors on a compact source sample. All definitions regarding the compactness of IFRSs are based on the apparent radio morphology. Defining a source as compact or extended is heavily dependent on the resolution of the observations and therefore a uniform classification is difficult. Using the criteria by \citet{2011A&A...531A..14Z} without an additional selection based on the apparent source size, we confirm the IFRS polarised source density found by \citet{2011A&A...526A...8M}, but not the high FP values found by these authors. This comparison has to be taken with care. The study by \citet{2011A&A...526A...8M} only consists of three polarised sources and is therefore highly susceptible to biases due to the small size of the sample. In addition, their instrumental leakage properties are higher, leading to an additional selection bias towards sources with higher FP values. We are also not able to find a difference in the median FP for resolved or extended sources between our complete sample and the IFRS one. This might suggest that IFRSs are not different  from typical polarised radio AGN regarding their polarisation characteristics.
        
        We want to note that our source sample is only representative of a typical polarised radio AGN sample. Polarised source samples on the $\mu$Jy-level are strongly biased towards AGN, while the total power samples in the radio regime are mostly a mixture of star-formation-dominated galaxies and AGN. In addition, even when only investigating the AGN sample, certain types of AGN with different characteristics dominate polarised samples. \citet{2015ApJ...806...83O} and \citet{2021arXiv210702492B} showed that a marginal difference exists between the FP of FRI and FRII sources, and also between radio-loud and radio-quiet sources. \citet{2012arXiv1209.1438H} found a difference between the FP of SDSS galaxies and radio-loud QSOs. The QSOs were found to have FP values  reaching as high as 30\,\%, while the FPs of SDSS galaxies reached only 15\,\%. 
        
        Comparing our classification distribution of the host galaxies with the one from \citet{2015ApJ...806...83O}, we see a large difference. While our sample mainly consists of intermediate disc and star-forming disc galaxies, the sample of \citet{2015ApJ...806...83O} is dominated by spheroidal ones. Making the same comparison with the faint total power radio sky at longer wavelengths shows a distribution very similar to ours \citep{2022arXiv220104433M}. The main difference arises in the flux range that the different samples cover. While \citet{2015ApJ...806...83O} used the NVSS, which probes the bright sky up to several mJy, our sensitivities are higher  by more than an order of magnitude. Therefore, our source population is dominated by faint polarised sources in the $\mu$Jy regime. 
        
        Several studies showed a discrepancy between the bright and faint polarised sky in terms of the relation between FP and TP radio emission (see \citet{2021arXiv210702492B} and references therein). This relation has been found to be larger for the bright sky. If the polarised source samples of the bright sky are indeed dominated by spheroidal galaxies, which are known to more often host the slightly stronger polarised FRI-type sources and exhibit less direct radio emission from star formation, we would expect the median FP of the bright sky to be lower than that from the spiral-galaxy- and QSO-dominated faint sky. Whether the bright and faint polarised sky are intrinsically different or this effect is a general selection bias remains unclear and further studies are needed. 
        
        Selection bias in the present study could originate from the limiting sensitivity of allWISE and SDSS. We base our analysis on the complementary optical and IR data. Spheroidal galaxies are less likely to be detected than brighter disc galaxies, meaning that counterparts for sources with late-type hosts are easier to detect. The fraction of sources classified as early-type by \citet{2015ApJ...806...83O} of more than 80\,\% is much larger than our fraction of 14.8\,\%. The total number of polarised sources we find with IR counterparts is 967 while the total number of polarised sources is 1170. If all remaining polarised sources without an IR counterpart had an early-type host galaxy, the fraction would only rise to 29.6\,\%, which is still very different from the percentage found by \citet{2015ApJ...806...83O}. We are therefore unable to completely exclude a bias towards disc-type galaxies in our sample, but this cannot explain the large differences we see.
        
        \section{Summary and outlook}
        \label{sect_summary}
        
        The Apertif system on the WSRT telescope is one of the two radio interferometers equipped with phased-array-feed technology allowing wide observations of the polarised radio sky in the centimetre wavelength regime. Using publicly available Science Verification Campaign data, we developed a semi-automatic pipeline to analyse the polarisation data products. We automatically mosaicked polarisation Stokes-Q and -U frequency cubes and performed RM synthesis. We generated PI, RM and FP maps from the resulting Faraday cubes. We used automatic source finding routines to generate polarised source catalogues, which are used to identify IR and optical counterparts of the polarised sources using the allWISE  and SDSS databases.
        
        We quantify the polarisation performance of the Apertif system by inspecting the polarisation leakage. We find that up to a primary beam response of 30\,\%, which is well within the overlap regions of individual beams for the mosaics, a maximum of only 1\,\% of the FP originates from leakage. Therefore, all sources with FPs of lower than 1\,\% were excluded from any further analysis. Additional performance checks were conducted by comparing the PI, FP, and RM values for sources with NVSS counterparts. We find only minimal differences between the NVSS values and ours, which we attribute to the factor-two lower resolution of the NVSS data.
        
        We identify 1170 polarised sources within 56.38\,deg$^2$, which results in a polarised source density of 21/deg$^2$. This is comparable to literature values at a similar wavelength and sensitivity. We estimated that at least two-thirds of our sources can be used to perform RM gridding, which is  more than in previous studies of the Milky Way and Large Magellanic Cloud by a factor of at least five. We show that our derived RMs originate from spatial variations in the magnetic field of the Milky Way foreground and are not the result of measurement uncertainties.
        
        For all of the sources of our sample with an IR counterpart (82.6\,\% of the total sample), we find that the total radio emission is dominated by the AGN activity. We find a subsample consisting of the sources with distance estimates  (27.3\,\%) to be mostly radio-loud (79\,\%) and of FRII-type (87\,\%). In contrast to the findings of an analysis of the bright sky using the NVSS, our source sample with IR counterparts ---surveying the faint sky--- is dominated by late-type galaxies instead of early-type ones. We are not able to explain this difference by selection biases. An intrinsic difference in the type of host galaxy for the bright and faint polarised sky might therefore be possible. 
        
        We notice an $\approx20\,\%$ higher FP for resolved sources than for unresolved sources. Reasons for this can be either beam depolarisation and/or intrinsic to the sources themselves. We did not observe a correlation between the FP and the star-formation density of the host galaxies.
        
        Although we have now doubled the survey area of polarised radio surveys at 1.4\,GHz in the $\mu$Jy regime, scientific analyses are still limited by the small number of detected polarised sources. The Apertif Wide Extragalactic Survey (AWES) will again extend this database to several tens of thousands of polarised sources within an area of 2000 deg$^2$ in the northern hemisphere. Complementary data from the southern hemisphere will be available from the Polarisation Sky Survey of the Universe's Magnetism (POSSUM) survey conducted by the Australian Square Kilometre Array Pathfinder (ASKAP). The combination of Apertif and ASKAP data with surveys at different wavelengths such as the LOFAR Two-metre Sky Survey (LoTSS), the LOFAR LBA Sky Survey (LoLSS), and the VLA Sky Survey (VLASS) will enhance the capability of scientific analysis to explore polarisation spectra and therefore the magnetic field morphology in radio sources.
        
        \begin{acknowledgements}
        This research made use of Montage. It is funded by the National Science Foundation under Grant Number ACI-1440620, and was previously funded by the National Aeronautics and Space Administration's Earth Science Technology Office, Computation Technologies Project, under Cooperative Agreement Number NCC5-626 between NASA and the California Institute of Technology. The Montage distribution includes an adaptation of the MOPEX algorithm developed at the Spitzer Science Center. BA and AB acknowledge funding from the German Science Foundation DFG, within the Collaborative Research Center SFB1491 ''Cosmic Interacting Matters - From Source to Signal''. EAKA is supported by the WISE research programme, which is financed by the Dutch Research Council (NWO). KMH acknowledges funding from the State Agency for Research of the Spanish Ministry of Science, Innovation and Universities through the ''Center of Excellence Severo Ochoa'' awarded to the Instituto de Astrof\'isica de Andaluc\'ia (SEV-2017-0709); from grant RTI2018-096228-B-C31 (Ministry of Science, Innovation and Universities / State Agency for Research / European Regional Development Funds, European Union); and from the coordination of the participation in SKA-SPAIN, funded by the Ministry of Science and innovation (MICIN). JMvdH and KMH acknowledge funding from the European Research Council under the European Union’s Seventh Framework Programme (FP/2007-2013)/ERC Grant Agreement No. 291531 (‘HIStoryNU’). JvL acknowledges funding from the European Research Council under the European Union’s Seventh Framework Programme (FP/2007-2013)/ERC Grant Agreement n. 617199 (`ALERT'), and from Vici research programme `ARGO' with project number 639.043.815, financed by the Dutch Research Council (NWO). DV acknowledges support from the Netherlands eScience Center (NLeSC) under grant ASDI.15.406. RS acknowledges funding from the European Research Council under the European Union's Seventh Framework Programme (FP/2007-2013)/ERC Grant Agreement No. 617199. LCO acknowledges funding from the European Research Council under the European Union's Seventh Framework Programme (FP/2007-2013)/ERC Grant Agreement No. 617199.
        \end{acknowledgements}
        
        \bibliographystyle{aa}
        \bibliography{bibexport}
        
        %\clearpage
        
        \begin{appendix}
                
                \section{SVC polarised intensity images}
                \label{appendix:piimages}
                
                PI images of SVC-fields 1, 2, 4, and 5 are shown in the following while SVC field 3 is shown in Fig. \ref{image_PI_SVC}. The greyscale represents the PI in all images. Compound beams missing in the data are given in the caption of each image together with the final synthesised beam size.
                
                \begin{figure*}
                        \resizebox{\hsize}{!}{\includegraphics{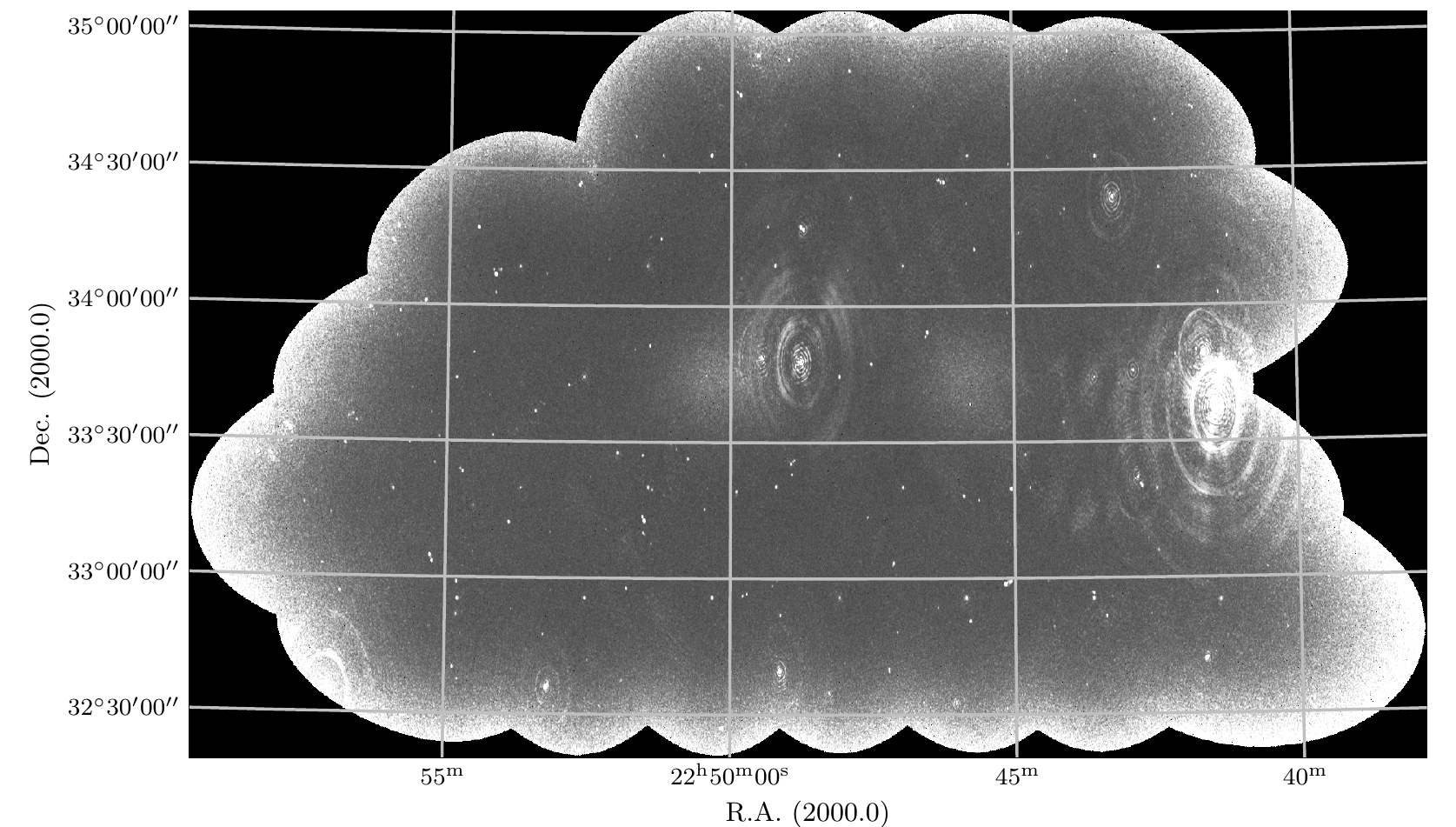}}
                        \caption{Polarised intensity image of the field S2248+33 (SVC1). The noise in emission-free regions is 14\,$\upmu$Jy/beam. The synthesised beam size is $32.1''\times18.0''$. Beams 16, 18, and 31 to 39 are missing.}
                        \label{image_PI_SVC1}
                \end{figure*}
                
                \begin{figure*}
                        \resizebox{\hsize}{!}{\includegraphics{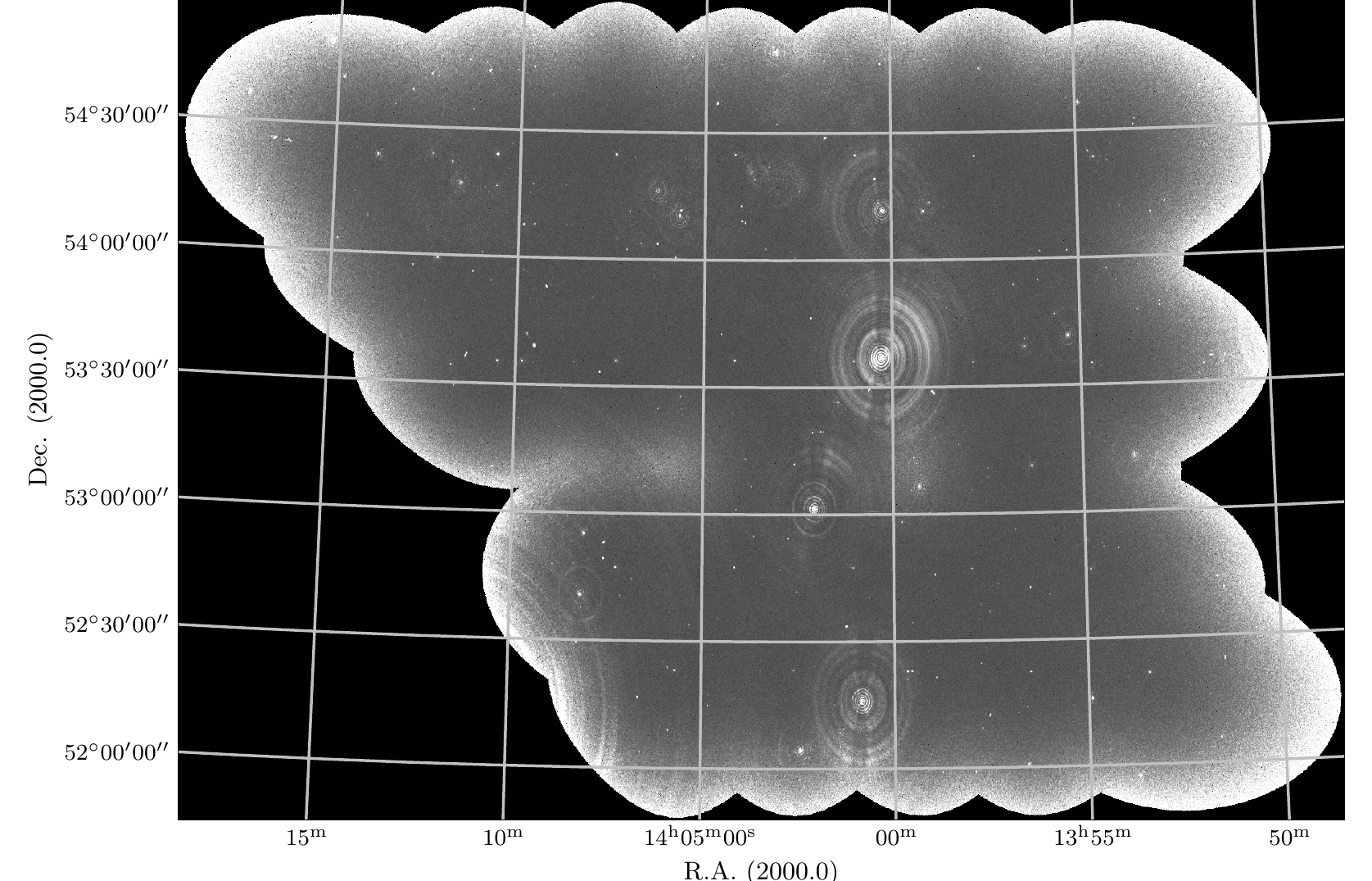}}
                        \caption{Polarised intensity image of the field M1403+53 (SVC2). The noise in emission-free regions is 13\,$\upmu$Jy/beam. The synthesised beam size is $19.3''\times14.6''$. Beams 6, 7, 13, 14, 16, 18 and 20 are missing.}
                        \label{image_PI_SVC2}
                \end{figure*}
                
                \begin{figure*}
                        \resizebox{\hsize}{!}{\includegraphics{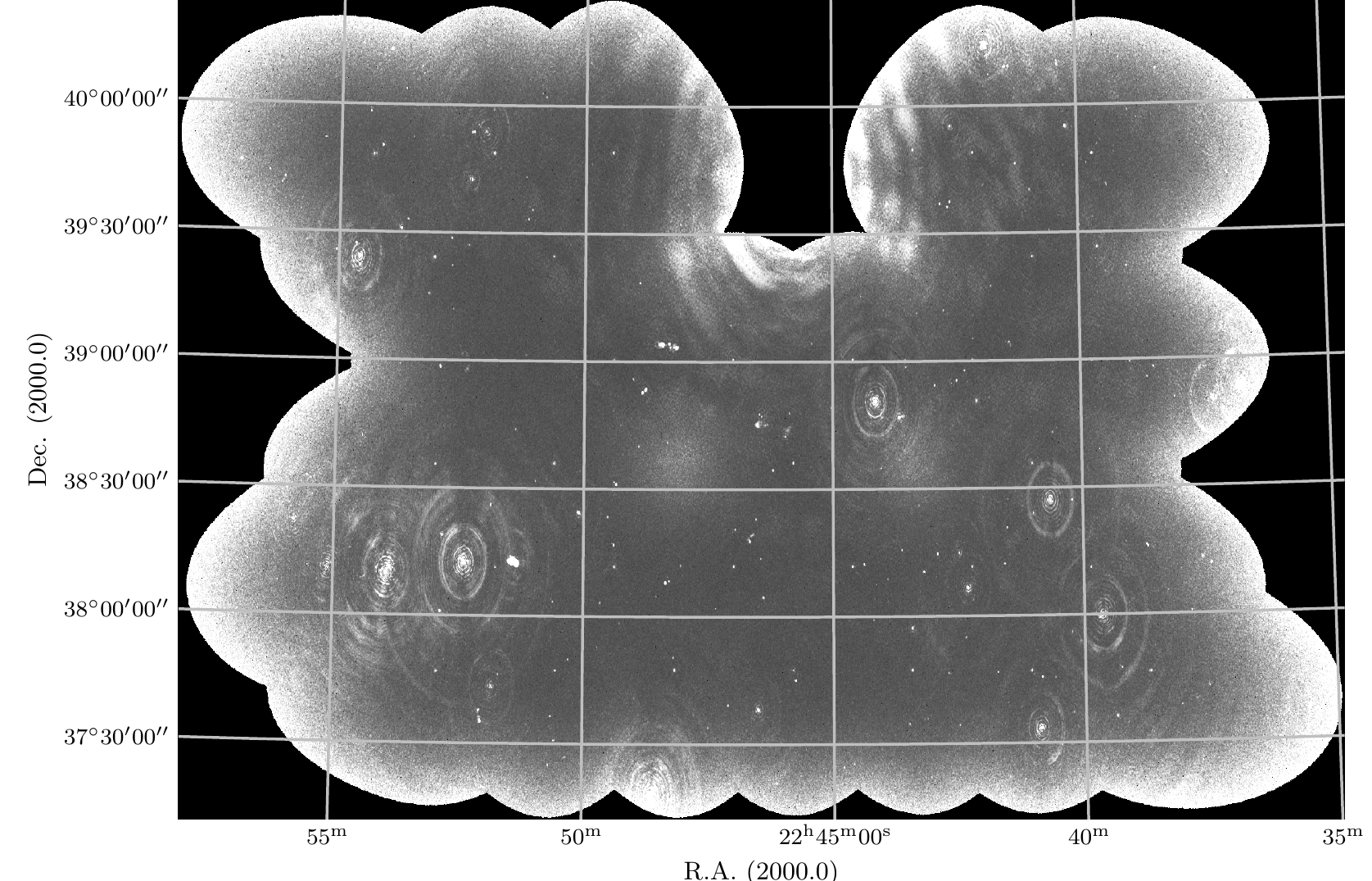}}
                        \caption{Polarised intensity image of the field S2246+38 (SVC4). The noise in emission-free regions is 16\,$\upmu$Jy/beam. The synthesised beam size is $24.8''\times14.8''$. Beams 16, 18, 28, 29, 30, 35 and 36 are missing.}
                        \label{image_PI_SVC4}
                \end{figure*}
                
                \begin{figure*}
                        \resizebox{\hsize}{!}{\includegraphics{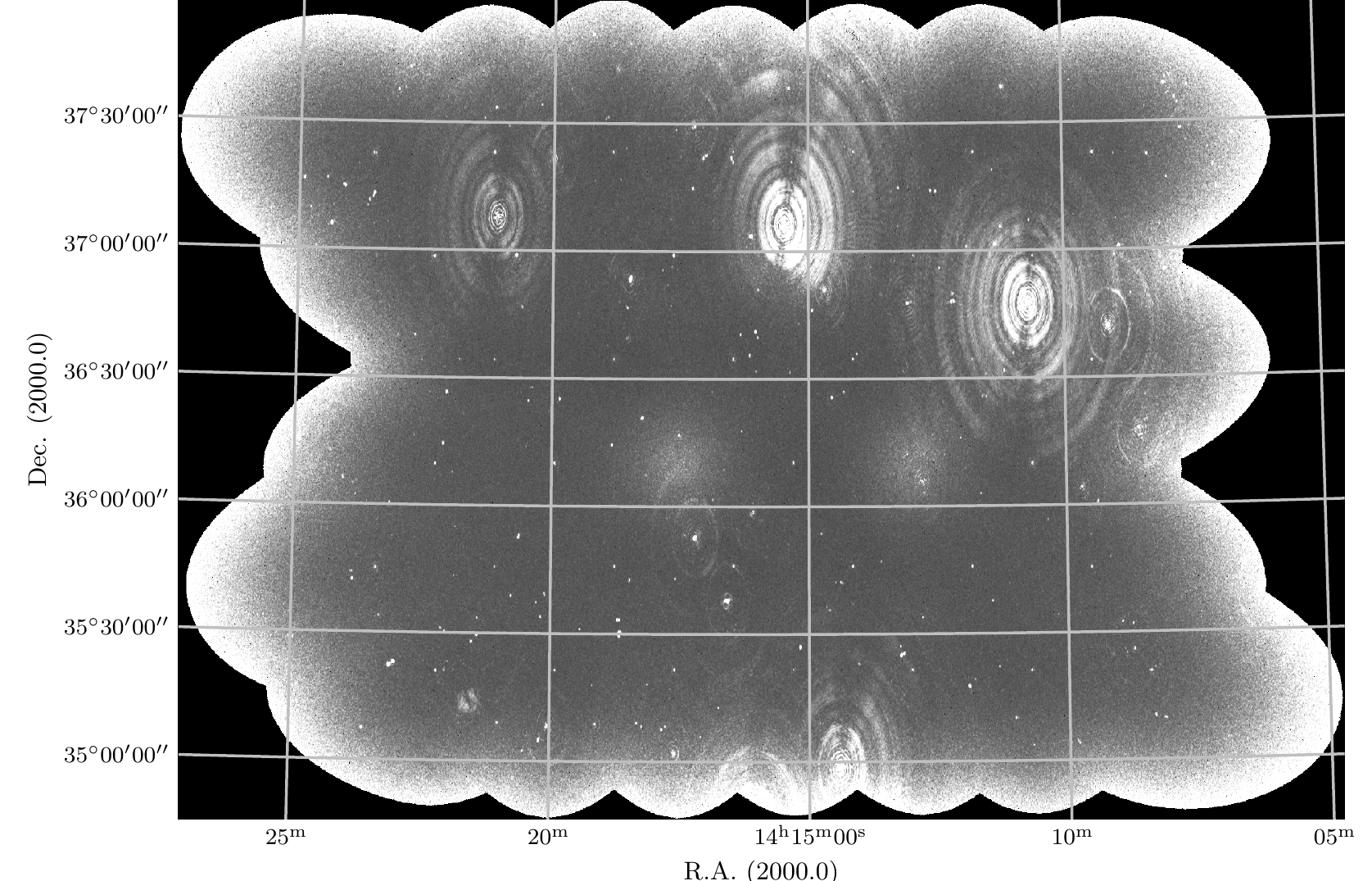}}
                        \caption{Polarised intensity image of the field S1415+36 (SVC5). The noise in emission-free regions is 16\,$\upmu$Jy/beam. The synthesised beam size is $31.2''\times17.7''$. Beams 16, 18 and 29 are missing.}
                        \label{image_PI_SVC5}
                \end{figure*}
                
%               \FloatBarrier
                
                \section{Source catalogue}
                \label{appendix_source_catalogue}
                
                Our polarised source catalogues are released as ascii files and are available via the ASTRON Virtual Observatotry\footnote{\url{https://vo.astron.nl/}}. In the following, we describe the entries for each column. Each row in the catalogue represents one source component, and so multiple entries with the same source ID are multiple polarised components of one source. All NVSS information is retrieved from \citet{2009ApJ...702.1230T}, WISE data from \citet{2014yCat.2328....0C}, and SDSS data from \citet{2020ApJS..249....3A}. NaN values are inserted where no information is available, which is usually the case for missing counterparts in the AllWISE or SDSS databases.
                
                \begin{description}
                        \item \textit{ID:} Source identifier concatenated from a prefix and the right ascension and declination of the central position of the source in degrees.
                        \item \textit{RA:} Right Ascension of the central position of the source in degrees.
                        \item \textit{RA\_err:} Uncertainty of the right Ascension of the central position of the source in degrees.
                        \item \textit{DEC:} Declination of the central position of the source in degrees.
                        \item \textit{DEC\_err:} Uncertainty of the declination of the central position of the source in degrees.
                        \item \textit{PI:} Integrated PI of the source in units of Jy.
                        \item \textit{PI\_err:} Uncertainty of the integrated PI of the source in units of Jy.
                        \item \textit{S\_Code:} S for unresolved sources, E for resolved sources
                        \item \textit{RA\_Comp:} Right ascension of the position of the source component in degrees.
                        \item \textit{RA\_Comp\_err:} Uncertainty of the right ascension of the position of the source component in degrees.
                        \item \textit{DEC\_Comp:} Declination of the position of the source component in degrees.
                        \item \textit{DEC\_Comp\_err:} Uncertainty of the declination of the position of the source component in degrees.
                        \item \textit{PI\_Comp\_peak:} Peak polarised emission of the source component in units of Jy.
                        \item \textit{PI\_Comp\_peak\_err:} Uncertainty of the peak polarised emission of the source component in units of Jy.
                        \item \textit{PI\_rms:} Standard deviation at the peak position of the source component in units of Jy/beam.
                        \item \textit{RM\_Comp:} RM at the peak position of the source component in units of rad/m$^2$.
                        \item \textit{RM\_Comp\_err:} Uncertainty of the RM at the peak position of the source component in units of rad/m$^2$.
                        \item \textit{TP\_ID:} TP source identifier. For internal usage only. Was used for cross-matching.
                        \item \textit{TP:} Radio continuum flux of the cross-matched TP counterpart in units of Jy.
                        \item \textit{TP\_err:} Uncertainty of the radio continuum flux of the cross-matched TP counterpart in units of Jy.
                        \item \textit{TP\_RA:} Right ascension of the central position of the TP counterpart in units of degrees.
                        \item \textit{TP\_RA\_err:} Uncertainty of the right ascension of the central position of the TP counterpart in units of degrees.
                        \item \textit{TP\_DEC:} Declination of the central position of the TP counterpart in units of degrees.
                        \item \textit{TP\_DEC\_err:} Uncertainty of the declination of the central position of the TP counterpart in units of degrees.
                        \item \textit{FP:} FP of the source derived from the integrated PI and TP fluxes.
                        \item \textit{FP\_err:} Uncertainty of the FP of the source derived from the integrated PI and TP fluxes.
                        \item \textit{NVSS\_I:} TP flux of a possible NVSS counterpart in units of Jy.
                        \item \textit{NVSS\_I\_err:} Uncertainty of the TP flux of a possible NVSS counterpart in units of Jy.
                        \item \textit{NVSS\_PI:} PI flux of a possible NVSS counterpart in units of Jy.
                        \item \textit{NVSS\_PI\_err:} Uncertainty of the PI flux of a possible NVSS counterpart in units of Jy.
                        \item \textit{NVSS\_FP:} FP of a possible NVSS counterpart.
                        \item \textit{NVSS\_FP\_err:} Uncertainty of the FP of a possible NVSS counterpart.
                        \item \textit{NVSS\_RM:} RM of a possible NVSS counterpart in units of rad/m$^2$.
                        \item \textit{NVSS\_RM\_err:} Uncertainty of the RM of a possible NVSS counterpart in units of rad/m$^2$.
                        \item \textit{NVSS\_RA:} Right ascension of a possible NVSS counterpart in units of degrees.
                        \item \textit{NVSS\_RA\_err:} Uncertainty of the right ascension of a possible NVSS counterpart in units of degrees.
                        \item \textit{NVSS\_DEC:} Declination of a possible NVSS counterpart in units of degrees.
                        \item \textit{NVSS\_DEC\_err:} Uncertainty of the declination of a possible NVSS counterpart in units of degrees.
                        \item \textit{WISE\_ID:} AllWISE ID of a possible counterpart.
                        \item \textit{WISE\_RA:} Right ascension of a possible AllWISE counterpart in units of degrees.
                        \item \textit{WISE\_RA\_err:} Uncertainty of the right ascension of a possible AllWISE counterpart in units of degrees.
                        \item \textit{WISE\_DEC:} Declination of a possible AllWISE counterpart in units of degrees.
                        \item \textit{WISE\_DEC\_err:} Uncertainty of the declination of a possible AllWISE counterpart in units of degrees.
                        \item \textit{WISE\_Flux\_3.4:} Brightness of a possible AllWISE counterpart in units of mag at 3.4$\upmu$m.
                        \item \textit{WISE\_Flux\_3.4\_err:} Uncertainty of the brightness of a possible AllWISE counterpart in units of mag at 3.4$\upmu$m.
                        \item \textit{WISE\_Flux\_3.4\_snr:} S/N of a possible AllWISE counterpart in units of mag at 3.4$\upmu$m.
                        \item \textit{WISE\_Flux\_4.6:} Brightness of a possible AllWISE counterpart in units of mag at 4.6$\upmu$m.
                        \item \textit{WISE\_Flux\_4.6\_err:} Uncertainty of the brightness of a possible AllWISE counterpart in units of mag at 4.6$\upmu$m.
                        \item \textit{WISE\_Flux\_4.6\_snr:} S/N of a possible AllWISE counterpart in units of mag at 4.6$\upmu$m.
                        \item \textit{WISE\_Flux\_12:} Brightness of a possible AllWISE counterpart in units of mag at 12$\upmu$m.
                        \item \textit{WISE\_Flux\_12\_err:} Uncertainty of the brightness of a possible AllWISE counterpart in units of mag at 12$\upmu$m.
                        \item \textit{WISE\_Flux\_12\_snr:} S/N of a possible AllWISE counterpart in units of mag at 12$\upmu$m.
                        \item \textit{WISE\_Flux\_22:} Brightness of a possible AllWISE counterpart in units of mag at 22$\upmu$m.
                        \item \textit{WISE\_Flux\_22\_err:} Uncertainty of the brightness of a possible AllWISE counterpart in units of mag at 22$\upmu$m.
                        \item \textit{WISE\_Flux\_22\_snr:} S/N of a possible AllWISE counterpart in units of mag at 22$\upmu$m.
                        \item \textit{SDSS\_ID:} SDSS DR16 ID of a possible counterpart.
                        \item \textit{SDSS\_RA:} Right ascension of a possible SDSS DR16 counterpart in units of degrees.
                        \item \textit{SDSS\_DEC:} Declination of a possible SDSS DR16 counterpart in units of degrees.
                        \item \textit{SDSS\_Flux\_U:} Brightness of a possible SDSS DR16 counterpart in units of mag in the U filter.
                        \item \textit{SDSS\_Flux\_U\_err:} Uncertainty of the brightness of a possible SDSS DR16 counterpart in units of mag in the U filter.
                        \item \textit{SDSS\_Flux\_G:} Brightness of a possible SDSS DR16 counterpart in units of mag in the G filter.
                        \item \textit{SDSS\_Flux\_G\_err:} Uncertainty of the brightness of a possible SDSS DR16 counterpart in units of mag in the G filter.
                        \item \textit{SDSS\_Flux\_R:} Brightness of a possible SDSS DR16 counterpart in units of mag in the R filter.
                        \item \textit{SDSS\_Flux\_R\_err:} Uncertainty of the brightness of a possible SDSS DR16 counterpart in units of mag in the R filter.
                        \item \textit{SDSS\_Flux\_I:} Brightness of a possible SDSS DR16 counterpart in units of mag in the I filter.
                        \item \textit{SDSS\_Flux\_I\_err:} Uncertainty of the brightness of a possible SDSS DR16 counterpart in units of mag in the I filter.
                        \item \textit{SDSS\_Flux\_Z:} Brightness of a possible SDSS DR16 counterpart in units of mag in the Z filter.
                        \item \textit{SDSS\_Flux\_Z\_err:} Uncertainty of the brightness of a possible SDSS DR16 counterpart in units of mag in the Z filter.
                        \item \textit{SDSS\_z:} Photometric or if available sprectroscopic redshift of a possible SDSS DR16 counterpart.
                        \item \textit{SDSS\_z\_err:} Uncertainty of the photometric or if available sprectroscopic redshift of a possible SDSS DR16 counterpart.
                \end{description}
                
        \end{appendix}
        
\end{document}